\documentclass[twocolumn,showpacs,preprintnumbers,amsmath,amssymb,pre]{revtex4}

\usepackage{graphicx}% Include figure files
\usepackage{dcolumn}% Align table columns on decimal point
\usepackage{bm}% bold math
\usepackage{enumerate}
\usepackage{amsmath}
\usepackage{color}

\begin{document}

\title{Model of anomalous diffusion--absorption process in a system consisting of two different media separated by a thin membrane}

\author{Tadeusz Koszto{\l}owicz}
 \email{tadeusz.kosztolowicz@ujk.edu.pl}
 \affiliation{Institute of Physics, Jan Kochanowski University,\\
         ul. \'Swi\c{e}tokrzyska 15, 25-406 Kielce, Poland}

\date{\today}

\begin{abstract}
We present the model of a diffusion-absorption process in a system which consists of two media separated by a thin partially permeable membrane. The kind of diffusion as well as the parameters of the process may be different in both media. Based on a simply model of particle's random walk in a membrane system we derive the Green's functions, then we find the boundary conditions at the membrane. One of the boundary conditions are rather complicated and takes a relatively simple form in terms of the Laplace transform. Assuming that particles diffuse independently of one another, the obtained boundary conditions can be used to solve to differential or differential-integral equations describing the processes in multi-layered systems for any initial condition. We consider normal diffusion, subdiffusion and slow subdiffusion processes, and we also suggest how superdiffusion could be included in this model. The presented method provides the functions in terms of the Laplace transform and some useful methods of calculation the inverse Laplace transform are shown.
\end{abstract}

                            
\maketitle

\section{Introduction\label{SecI}}

In many systems we can meet in biology, medicine, physics and engineering sciences various kinds of diffusion occur in a system composed of different media separated by a thin membrane \cite{hobbie,luckey,hsieh}. We mention here diffusion of various substances through the skin \cite{schumm}, in the brain \cite{zhan,tao,linninger}, and between blood and a cell \cite{kim}; a list of similar examples can be significantly extended. Diffusing particles can be also absorbed, with some probability, in the media. There may be a different kinds of anomalous diffusion in each medium, which are described by differential or differential-integral equations. To solve the equations, two boundary conditions at the membrane are needed. 
However, until now, various boundary conditions which are not equivalent to one another have been assumed at the membrane, see for example \cite{zhang,grebenkov,kdl,tk,tk1,korabel1,korabel2,korabel3,singh,kim1,ash,huang,adrover,abdekhodaie,taveira,cabrera1,gp}. 
In many papers, the boundary conditions with respect to normal diffusion or subdiffusion have been just assumed or derived by means of phenomenological models. In this paper we derive the Green's functions by means of the particle's random walk with absorption model in a system with a thin membrane. Knowing the Green's functions we derive the boundary conditions at the thin membrane. Similar procedure of deriving boundary condition at fully absorbing or fully reflecting wall was used by Chandrasekhar \cite{chandra}. Some aspects of discrete random walk model presented in this paper and some special cases of diffusion in a two--layered system were already published, namely subdiffusion without absorption in a system with a double--sided partially permeable membrane \cite{tk,tk1,tk2} and subdiffusion with absorption in a system in which the boundary between the media is fully permeable for diffusing particles \cite{tk3}.

\begin{figure}[h]
        \includegraphics[scale=0.4]{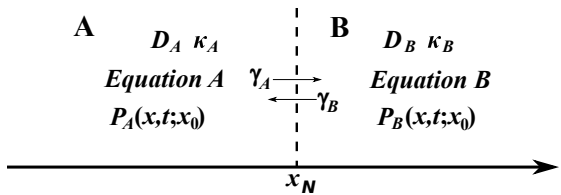}
				\caption{The system which consists of two media $A$ and $B$ separated by a thin membrane located at $x_N$, $P_A$ and $P_B$ denote the Green's functions, $D_A$, $D_B$ are the generalized diffusion coefficients, $\kappa_A$ and $\kappa_B$ denote absorption coefficients, $\gamma_A$ and $\gamma_B$ are membrane permeability coefficients. The diffusion--absorption processes are described by some differential or differential--integral equations defined separately in the parts $A$ and $B$, the boundary conditions at the membrane are to be determined. \label{Fig1}}
\end{figure}

The main aim of this paper is to present a universal model that leads to general diffusion--absorption equations, Green's functions and boundary conditions at a thin membrane for a system consisting of two media $A$ and $B$ separated by a thin membrane, in each medium there may be a different type of diffusion, see Fig. \ref{Fig1}. The diffusion and absorption parameters, that can be different in both media, as well as the membrane permeability parameters are assumed to be constant. The universality of this model lies in the fact that the kind of diffusion in a medium is determined by one function alone, hereinafter referred to as $v$. This function controls of time which is needed to take the diffusing particle's next step. The membrane permeability parameters are determined by the probabilities of the single particle's passing through the membrane; the probabilities can be calculated using a phenomenological model. The thin membrane represents any obstacle which can stop diffusing particle with some probability. We also consider the process in a system with a one-sided fully permeable membrane, this case is qualitatively different from the case of a double--sided partially permeable membrane. The boundary conditions at the membrane depend on two functions $v_A$ and $v_B$ defining the type of diffusion in both media, the conditions are `adapted' to the kind of diffusion processes occurring on both sides of the membrane. 

We consider diffusion--absorption processes in a system which is homogeneous in a plane perpendicular to the $x$ axis, thus the system is effectively one-dimensional. The considered system is shown schematically in Fig. \ref{Fig1}.
The Green's function $P(x,t;x_0)$ is interpreted as the probability density of finding a diffusing particle at the point $x$ at time $t$, $x_0$ is the initial position of the particle. This function is also defined as the solution to the diffusion equation for the initial condition expressed by the delta-Dirac function, $P(x,0;x_0)=\delta(x-x_0)$. Knowing Green's functions for both regions we can derive the boundary conditions at the membrane. Assuming that diffusing particles move independently of one another, the obtained boundary conditions can be used for any initial concentration. 
The normal diffusion, `classical' subdiffusion and slow subdiffusion processes, all with absorption, are included in the model. In the Final Remarks we also suggest the method of involving superdiffusion into the model. 

We start our considerations with the model of random walk in a system in which time and spatial variables are discrete. Then, we move to continuous variables. This method is slightly different from the `classical' Continuous Time Random Walk (CTRW) method \cite{montroll65,mk,ks}. Namely, in the CTRW method the time which is needed to take the particle's next step $\tau$ and the length of the particle's jump $\epsilon$ are both random variables, while in the method presented in this paper $\tau$ is a random variable whereas $\epsilon$ is a parameter. The motivation to involve discrete model into considerations is that the difference equations describing random walk in the membrane system are solvable. These equations has also a very simple interpretation.  
Parameters describing random walk in a discrete system, like probability of particle's absorbing and probability of stopping a particle by the membrane should be redefined in a system with continuous variables. This is one of the main problems how to define the parameters in the system with continuous variables and to derive relations linking these parameters with the probabilities specified in the discrete system.

The organization of the paper is as follows. In Sec. \ref{SecII} we present the general procedure of deriving normal diffusion, subdiffusion, and slow subdiffusion--absorption equations and Green's functions for homogeneous system. The procedure is based on the particle's random walk model in a system with discrete time and spatial variables. Finding the generating function for the difference equation describing the particle's random walk we move from discrete to continuous variables. The random walk model with absorption in a system which consists of two different media separated by a thin membrane is considered in Sec. \ref{SecIII}. We derive Green's functions and boundary conditions in terms of the Laplace transform. The model presented in this paper provides the results that the boundary conditions at the border between media can depend on in which medium the particle is initially located. This fact causes that the procedure of solving the system of diffusion-absorption equations for arbitrarily chosen initial condition is somewhat complicated. This procedure is presented in Sec. \ref{SecIV}. As an example, in Sec. \ref{SecV} we consider diffusion in a system in which one-sided fully permeable membrane separates subdiffusive medium $A$ and medium $B$ in which subdiffusion or slow subdiffusion with absorption occurs. Final remarks are presented in Sec. \ref{SecVI}. Since calculations of the inverse Laplace transforms appear to be difficult, in the Appendix I we show some methods of such calculations useful for the functions presented in this paper. Details of some calculations are presented in the Appendix II.

\section{Random walk model of diffusion with absorption in a homogeneous system\label{SecII}}

Diffusion with absorption $X+Y\rightarrow Y$, where $X$ represents a diffusing particle and $Y$ is an `absorbing point', has been modelled using a discrete random walk on a lattice model \cite{rod,abr,koszt2014}. 
Here we consider diffusion with absorption described by the following difference equation
\begin{eqnarray}\label{eq1}
  P_{n+1}(m;m_0)=\frac{1}{2}P_{n}(m-1;m_0)+\frac{1}{2}P_{n}(m+1;m_0)\\
	-RP_{n}(m;m_0)\;,\nonumber
\end{eqnarray}
where $P_{n}(m;m_0)$ is the probability of finding a particle at site $m$ after $n$ steps, $m_0$ denotes the initial position of the particle, $P_{0}(m;m_0)=\delta_{m,m_0}$, $R$ is the probability of absorption.
In further considerations, we will use the generating function defined as
\begin{equation}\label{eq2}
S(m,z;m_0)=\sum_{n=0}^\infty z^n P_{n}(m;m_0)\; . 
\end{equation}
To move from discrete to continuous time we use the standard formula \cite{montroll65}
\begin{equation}\label{eq3}
  P(m,t;m_0)=\sum_{n=0}^{\infty}P_n(m,m_0)\Phi_n(t)\;,
\end{equation}
where $\Phi_n(t)$ is the probability that the particle takes $n$ steps over a time interval $[0,t]$. 
The function is the convolution
\begin{eqnarray}\label{eq4}
\Phi_n(t)=\int_0^{t_2}dt_1\int_0^{t_3}dt_2\ldots\int_0^{t_n}dt_{n-1}\omega(t_1)\omega(t_2-t_1)\\
\times\ldots\omega(t_n-t_{n-1})U(t-t_n),\nonumber
\end{eqnarray}
where $\omega(t)$ is the probability density of time which is needed for the particle to take its next step, $U(t)=1-\int_0^t \omega(t')dt'$ is the probability that the particle has not performed any step over a time interval $[0,t]$.
It is convenient to carry out further calculations in terms of the Laplace transform $\mathcal{L}[f(t)]\equiv \hat{f}(s)=\int_0^\infty {\rm e}^{-st}f(t)dt$ due to the relation $\mathcal{L}\bigg[\int_0^t f(t')g(t-t')dt'\bigg]=\hat{f}(s)\hat{g}(s)$. 
The Laplace transform of $\Phi_n(t)$ reads 
\begin{equation}\label{eq5}
\hat{\Phi}_n(s)=\hat{U}(s)\hat{\omega}^n(s),
\end{equation}
where 
\begin{equation}\label{eq6}
\hat{U}(s)=\frac{1-\hat{\omega}(s)}{s}.
\end{equation}
From Eqs. (\ref{eq3}), (\ref{eq5}), and (\ref{eq6}) we get
\begin{eqnarray}\label{eq7}
  \hat{P}(m,s;m_0)=\frac{1-\hat{\omega}(s)}{s}S\left(m,\hat{\omega}(s);m_0\right)\;. 
\end{eqnarray}

Moving from discrete to continuous space variable we use the following relations
\begin{equation}\label{eq8}
x=\epsilon m\;,\;x_0=\epsilon m_0\;,
\end{equation}
and
\begin{equation}\label{eq9}
P(x,t;x_0)=\frac{P(m,t;m_0)}{\epsilon}\;.
\end{equation}
The parameter $\epsilon$ is the distance between neighbouring sites, which can be interpreted as a length of single particle jump. In the following we conduct the considerations in the limit of small $\epsilon$.

There is $\hat{\omega}(0)=1$ due to the normalization of the function $\omega(t)$. Within CTRW normal diffusion or subdiffusion is usually considered in the long time limit \cite{mk,ks}, which corresponds to the limit of small $s$, under assumption that $\hat{\omega}(s)=1-\mu s^\alpha$, $0<\alpha\leq 1$, $\mu$ is a positive parameter. This formula can be written in a more general form
\begin{equation}\label{eq10}
	\hat{\omega}(s)= 1-\mu v(s),
\end{equation}
where $v(s)$ is a function that $v(s)\rightarrow 0$ when $s\rightarrow 0$; the function $v(s)$ will define the kind of diffusion. We assume that the parameters occurring in $v(s)$ are dimensionless or their physical units are the same as the physical unit of $s$ which is the inverse of time unit. Then, $\mu$ is chosen in such a way that the last term in the right-hand side of Eq. (\ref{eq10}) is dimensionless. In the following considerations we will use the approximation of $\hat{\omega}(s)$ given by Eq. (\ref{eq10}), but we will find that the parameter $\epsilon$ controls $\mu$, and show that Eq. (\ref{eq10}) is valid in the limit of small parameter $\epsilon$ (which corresponds to the limit of small parameter $\mu$) for any positive $s$, see Eq. (\ref{eq20}) in Sec. \ref{SecIIB}. Thus, the long time approximation will not be needed.

\subsection{Laplace transform of Green's function\label{SecIIA}}

The generating function of Eq. (\ref{eq1}) is
\begin{equation}\label{eq11}
  S(m,z;m_0)=\frac{[\eta(z)]^{|m-m_0|}}{\sqrt{(1+zR)^2-z^2}},
\end{equation}
where
\begin{equation}\label{eq12}
  \eta(z)=\frac{1+zR-\sqrt{(1+zR)^2-z^2}}{z}.
\end{equation}
In the limit of small $s$ and $\epsilon$ we obtain from Eqs. (\ref{eq7})--(\ref{eq12})  
\begin{widetext}
\begin{eqnarray}\label{eq13}
\hat{P}(x,s;x_0)=\frac{\mu v(s)}{\epsilon s\sqrt{2R+R^2+2\mu (1-R-R^2)v(s)+\mu^2(R^2-1)v^2(s)}}\\
\times\left[\frac{1+R-\mu Rv(s)-\sqrt{2R+R^2+2\mu(1-R-R^2)v(s)+\mu^2(R^2-1)v^2(s)}}{1-\mu v(s)}\right]^{\frac{|x-x_0|}{\epsilon}}.\nonumber
\end{eqnarray}
\end{widetext}
Let us consider the conditions that will ensure that the function Eq. (\ref{eq13}) will not be equivalent to the zero function and will take finite values in the limit of small parameter $\epsilon$. For $R=0$ the above conditions are fulfilled only when $\epsilon\sim \sqrt{\mu}$. We define the generalized diffusion coefficient as 
\begin{equation}\label{eq14}
D=\frac{1}{2}\frac{\epsilon^2}{\mu}\;.
\end{equation}
For the case of $R\neq 0$ the above conditions and Eq. (\ref{eq14}) provide $\sqrt{2R+R^2}\sim\epsilon$. We assume that $\kappa=\sqrt{2R+R^2}/\epsilon$, where $\kappa$ is the absorption coefficient defined in the continuous system. The last relation provides $R=\sqrt{1+\epsilon^2\kappa^2}-1$. In the limit of small $\epsilon$ we get
\begin{equation}\label{eq15}
R=\frac{\kappa^2\epsilon^2}{2}.
\end{equation}

Taking into account Eqs. (\ref{eq13})--(\ref{eq15}), in the limit of small $\epsilon$ the Laplace transform of the Green's function reads
\begin{equation}\label{eq16}
\hat{P}(x,s;x_0)=\frac{v(s)}{2D s\sqrt{\kappa^2+\frac{v(s)}{D}}}\;{\rm e}^{-|x-x_0|\sqrt{\kappa^2+\frac{v(s)}{D}}}.
\end{equation}

\subsection{Diffusion equation\label{SecIIB}}

We derive the diffusion equation in terms of the Laplace transform starting from Eq. (\ref{eq1}).
Combining Eqs. (\ref{eq1}), (\ref{eq2}), and (\ref{eq7}) we get
\begin{eqnarray}\label{eq17}
\frac{1}{z}\left[S(m,z;m_0)-P_0(m;m_0)\right]=\frac{1}{2}S(m-1,z;m_0)\\
+\frac{1}{2}S(m+1,z;m_0)-R S(m-1,z;m_0)\;.\nonumber
\end{eqnarray}
From Eqs. (\ref{eq7})--(\ref{eq9}), (\ref{eq17}) and the relation $\partial^2 f(x)/\partial x^2\approx [f(x+\epsilon)+f(x-\epsilon)-2f(x)]/\epsilon^2$ we obtain
\begin{eqnarray}\label{eq18}
s\hat{P}(x,s;x_0)-P(x,0;x_0)\\
=\frac{\epsilon^2 s\hat{\omega}(s)}{2(1-\hat{\omega}(s))}\left[\frac{\partial^2 \hat{P}(x,s;x_0)}{\partial x^2}-\kappa^2\hat{P}(x,s;x_0)\right]\;.\nonumber
\end{eqnarray} 
The Green's function Eq. (\ref{eq16}) fulfils Eq. (\ref{eq18}) only if
\begin{equation}\label{eq19}
\hat{\omega}(s)=\frac{1}{1+\frac{\epsilon^2 v(s)}{2D}}\;.
\end{equation}
In the limit of small $\epsilon$ Eq. (\ref{eq19}) can be approximated as 
\begin{equation}\label{eq20}
\hat{\omega}(s)= 1-\epsilon^2\frac{v(s)}{2D}
\end{equation}
for any positive $s$. 

Eqs. (\ref{eq18}) and (\ref{eq19}) provide
\begin{eqnarray}\label{eq21}
\frac{v(s)}{s}\left[s\hat{P}(x,s;x_0)-P(x,0;x_0)\right]\\=D\Bigg[\frac{\partial^2 \hat{P}(x,s;x_0)}{\partial x^2}
-\kappa^2\hat{P}(x,s;x_0)\Bigg].\nonumber
\end{eqnarray}
In the time domain the general form of the diffusion--absorption equation reads
\begin{eqnarray}\label{eq23}
\int_0^t F(t-t')\frac{\partial P(x,t';x_0)}{\partial t'}dt'\\
=D\Bigg[\frac{\partial^2 P(x,t;x_0)}{\partial x^2}
-\kappa^2 P(x,t;x_0)\Bigg],\nonumber
\end{eqnarray}
where
\begin{equation}\label{eq24}
F(t)=\mathcal{L}^{-1}\left[\frac{v(s)}{s}\right].
\end{equation}
Eq. (\ref{eq23}) represents Eq. (\ref{eq21}) in the time domain only if the inverse Laplace transform of $v(s)/s$ exists.
The diffusive flux $J$ is defined in terms of the Laplace transform as follows
\begin{equation}\label{eq22}
\hat{J}(x,s;x_0)=-D\frac{s}{v(s)}\frac{\partial \hat{P}(x,s;x_0)}{\partial x}.
\end{equation}
Combining Eq. (\ref{eq22}) with the Laplace transform of the continuity equation, $\partial P/\partial t=-\partial J/\partial x $, we get Eq. (\ref{eq21}) for diffusion without absorption, $\kappa=0$.

The probability $\mathcal{P}(t)$ that a diffusing particle has not been absorbed until time $t$ equals $\mathcal{P}(t)=1-\int_{-\infty}^\infty P(x,t;x_0)dx$. In term of the Laplace transform we obtain from Eq. (\ref{eq16}) $\hat{\mathcal{P}}(s)=\kappa^2/[s(\kappa^2+v(s)/D)]$. The probability depends on the kind of diffusion process. This fact imposes the following interpretation of the diffusion-absorption process. Absorption can be treated as an irreversible reaction $X+Y\rightarrow Y$, where $X$ represents a diffusing particle and $Y$ is an absorbing point. The absorbing points are assumed to be distributed homogeneously in the system, then the probability of reaction does not depend on the position of the particle. The absorption process of a diffusing particle consists of two stages. In the first stage, the particle $X$ can meet the point $Y$, with some probability, after its jump. If this event occurs, absorption of the particle may occur in the second stage. Since the occurrence of the first stage depends on the kind of diffusion, the probability that the particle still exists in the system at time $t$ also depends on the kind of diffusion. We note that this interpretation is not valid if the first stage occurs with the probability equals 1, then the `absorption process' is equivalent to the reaction $X\rightarrow\emptyset$. In this case the process cannot be described by Eq. (\ref{eq21}), see \cite{sss} and the discussion in \cite{kl}; this problem is discussed in more detail in Sec. \ref{SecVI} point 7. 

\subsection{Normal diffusion, subdiffusion and slow subdiffusion\label{SecIIC}}

We define normal diffusion, subdiffusion and slow subdiffusion (which is also called `ultraslow diffusion') by means of fractional moments of the function $\omega(t)$. The fractional moment of the order $\rho>0$ is defined as
$\left\langle \tau^\rho\right\rangle\equiv \int_0^\infty \tau^\rho \omega(\tau)d\tau$. 
The moment of fractional order $\rho$ can be obtained using the equation 
$\left\langle \tau^\rho\right\rangle =\left(-1/\Gamma(k-\rho)\right)^k\int_0^\infty ds\; s^{k-\rho-1}d^k\hat{\omega}(s)/ds^k$,
where $k$ is the smallest natural number such that $k>\rho$. 
For the moment of natural order $k$ this formula takes the form 
\begin{equation}\label{eq24b}
\left\langle \tau^k \right\rangle=(-1)^k \frac{d^k\hat{\omega}(s)}{ds^k}\Bigg|_{s=0}.
\end{equation} 

Diffusion is often characterized by temporal evolution of the mean square displacement of the particle $\left\langle (\Delta x)^2(t)\right\rangle\equiv\int_{-\infty}^\infty (x-x_0)^2 P(x,t;x_0)dx$ in the system without absorption, for $\kappa=0$. From Eq. (\ref{eq16}), in terms of the Laplace transform we get 
\begin{equation}\label{eq24a}
\left\langle (\Delta x)^2(t)\right\rangle=\mathcal{L}^{-1}\left[\frac{2D}{sv(s)}\right].
\end{equation}
Various kinds of anomalous diffusion and their characteristics based on Eq. (\ref{eq24a}) were considered in \cite{mj,chk}.

\subsubsection{Normal diffusion\label{SecIICA}}

Normal diffusion is defined as a process in which $\left\langle \tau\right\rangle=\int_0^\infty \tau\omega(\tau)d\tau<\infty$, then $v(s)=s$. In this case $\left\langle (\Delta x)^2(t)\right\rangle=2Dt$.
From Eq.~(\ref{eq21}) we get the normal diffusion--absorption equation
\begin{equation}\label{eq25}
  \frac{\partial P(x,t;x_0)}{\partial t}=D\left[\frac{\partial^2 P(x,t;x_0)}{\partial x^2}-\kappa^2 P(x,t;x_0)\right]\;.
\end{equation}

\subsubsection{Subdiffusion\label{SecIICB}}

In the case of `classical' subdiffusion there exists a parameter $\alpha$, $0<\alpha<1$, for which $\left\langle \tau^\rho \right\rangle=\infty$ for $\rho>\alpha$, and $\left\langle \tau^\rho\right\rangle<\infty$ for $\rho\leq\alpha$. In this case $v(s)=s^\alpha$ which provides $\left\langle (\Delta x)^2(t)\right\rangle=2Dt^\alpha/\Gamma(1+\alpha)$.
Due to Eq. (\ref{eq21}) and the formula $\mathcal{L}^{-1}[s^\alpha \hat{g}(s)]=d^\alpha g(t)/dt^\alpha$, $0<\alpha<1$, the `classical' subdiffusion--absorption equation reads 
\begin{equation}\label{eq26}
  \frac{\partial P(x,t;x_0)}{\partial t}=D\frac{\partial^{1-\alpha}}{\partial t^{1-\alpha}}\left[\frac{\partial^2 P(x,t;x_0)}{\partial x^2}-\kappa^2 P(x,t;x_0)\right],
\end{equation}
$0<\alpha<1$, where the Riemann--Liouville fractional time derivative is defined for $\beta>0$ as 
\begin{equation}\label{eq27}
\frac{d^\beta f(t)}{dt^\beta}=\frac{1}{\Gamma(k-\beta)}\int_0^t dt'(t-t')^{k-1-\beta}f(t'),
\end{equation} 
the integer number $k$ fulfils the relation $k-1<\beta\leq k$. Eq. (\ref{eq26}) was considered in many papers, see for example \cite{henry,yuste,seki1,seki2,mfh}. Three models of subdiffusion--reaction process which provide different fractional equations with linear reaction term and their Green's functions for a homogeneous system were considered in \cite{henry}. Eq. (\ref{eq26}) corresponds to the subdiffusion--reaction equation derived in the above cited paper for the model in which absorption of a particle occurs with a certain probability immediately after particle's jump. 

\subsubsection{Slow subdiffusion\label{SecIICC}}

Slow subdiffusion can be defined as a process for which $\left\langle \tau^\rho \right\rangle =\infty$ for $\rho>0$. This condition is fulfilled when $v(s)$ is a slowly varying function \cite{denisov}. A slowly varying function $f$ fulfils the condition $f(au)/f(u)\rightarrow 1$ when $u\rightarrow\infty$ for any $a>0$. In practice, this function is a combination of logarithm functions or it has a finite limit when $u\rightarrow\infty$. Due to the Tauberian theorem we get $\left\langle (\Delta x)^2(t)\right\rangle=2D/v(1/t)$. The slow subdiffusion--absorption equation depends on the detailed form of the slowly varying function $v(s)$. For $v(s)=1/{\rm ln}^{r}(1/s)$, $r>0$, we have 
\begin{eqnarray}\label{eq28}
  \frac{1}{\Gamma(r)}\int_0^t \mu(t-t',r)\frac{\partial P(x,t';x_0)}{\partial t'}dt'\\
  =D\frac{\partial^2 P(x,t;x_0)}{\partial x^2}-\kappa^2 P(x,t;x_0)\;,\nonumber
\end{eqnarray}
where $\mu(t,r)=\int_0^\infty d\zeta\frac{t^\zeta \zeta^{r}}{\Gamma(1+\zeta)}$ is the Volterra--type function \cite{ob}. The qualitative difference between the Green's functions for `classical' subdiffusion and slow subdiffusion for $\kappa=0$ is discussed in \cite{tk1}.

\section{Diffusion--absorption process in a system in which thin membrane separates two media\label{SecIII}}

Let the symbol $A$ denotes the region $(-\infty,x_N)$ and the symbol $B$ denotes the region $(x_N,\infty)$, $x_N$ is the position of the thin membrane. The symbols will be also assigned to the functions and parameters defined in these regions.
Typically, diffusion processes with absorption are described by various differential or differential--integral equations. To solve the equations, two boundary conditions should be given at the thin membrane. The boundary conditions depend on the kind of processes taking place in the parts $A$ and $B$ and on the parameters $\gamma_A$ and $\gamma_B$ controlling the permeability of the membrane, see Fig. \ref{Fig1}.	
			\begin{figure}[h]
        \includegraphics[scale=0.4]{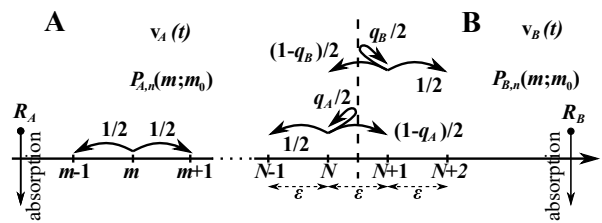}
				\caption{Random walk with absorption in a membrane system with discrete space variable $m$ and time $n$ which is the number of particle's steps, $R_A$ and $R_B$ are absorption probabilities. The functions $v_A$ and $v_B$ define the kinds of diffusion. The more detailed description is in the text. \label{Fig2}}
      \end{figure}
			
The idea of the method presented in this paper is as follows. Instead of the system presented in Fig. \ref{Fig1}, we consider the corresponding system with discrete variables shown in Fig. \ref{Fig2}.	
A particle performs its single jump to the neighbouring site only if the particle is not stopped by the membrane with a certain probability. The particle which tries to pass through the membrane moving from the $N$ to $N+1$ site can pass the membrane with probability $(1-q_A)/2$ or can be stopped by the membrane with probability $q_A/2$. When a particle is located at the $N+1$ site, then its jump to the $N$ site can be performed with probability $(1-q_B)/2$. The probability that a particle can be stopped by the membrane equals $q_B/2$. 
The difference equations describing the random walk in a membrane system with reactions are
\begin{eqnarray}\label{eq29}
 P_{A,n+1}(m;m_0)=\frac{1}{2}P_{A,n}(m-1;m_0)\\
+\frac{1}{2}P_{A,n}(m+1;m_0)-R_A P_{A,n}(m;m_0),\nonumber
\\ m\leq N-1,\nonumber
\end{eqnarray}
\begin{eqnarray}\label{eq30}
P_{A,n+1}(N;m_0)=\frac{1}{2}P_{A,n}(N-1;m_0)\\
+\frac{1-q_2}{2}P_{B,n}(N+1;m_0)+\frac{q_1}{2}P_{A,n}(N;m_0)\nonumber\\
-R_A P_{A,n}(N;m_0)\;,\nonumber
\end{eqnarray}
\begin{eqnarray}\label{eq31}
P_{B,n+1}(N+1;m_0)=\frac{1-q_1}{2}P_{A,n}(N;m_0)\\
+\frac{1}{2}P_{B,n}(N+2;m_0)+\frac{q_2}{2}P_{B,n}(N+1;m_0)\nonumber\\ 
-R_B P_{B,n}(N+1;m_0)\;,\nonumber
 \end{eqnarray}
\begin{eqnarray}\label{eq32}
P_{B,n+1}(m;m_0)=\frac{1}{2}P_{B,n}(m-1;m_0)\\
+\frac{1}{2}P_{B,n}(m+1;m_0)
-R_B P_{B,n}(m;m_0),\nonumber\\
 m\geq N+2.\nonumber
\end{eqnarray}
We assume that $m_0\leq N$, the initial conditions are 
\begin{equation}\label{eq33}
P_{A,0}(m;m_0)=\delta_{m,m_0}\;,P_{B,0}(m;m_0)=0.
\end{equation}

The generating functions are defined separately for the regions $A$ and $B$
\begin{equation}\label{eq34}
  S_i(m,z;m_0)=\sum_{n=0}^{\infty}z^nP_{i,n}(m,m_0)\;,
\end{equation}
$i=A,B$. 
After calculations, we obtain (the details of the calculations are presented in \cite{tk2})
\begin{eqnarray}\label{eq35} 
S_A(m,z;m_0)=\frac{[\eta_{A}(z)]^{|m-m_0|}}{\sqrt{(1+zR_A)^2-z^2}}\\
+\Lambda_A(z)\frac{[\eta_{A}(z)]^{2N-m-m_0}}{\sqrt{(1+zR_A)^2-z^2}}\;,\nonumber
\end{eqnarray}
\begin{equation}\label{eq36} 
  S_B(m,z;m_0)=\frac{[\eta_{A}(z)]^{N-m_0}[\eta_{B}(z)]^{m-N-1}}{\sqrt{(1+zR_B)^2-z^2}}\Lambda_B(z)\;,
\end{equation}
where
\begin{equation}\label{eq37}
\Lambda_A(z)=\frac{\Big(\frac{1}{\eta_{B}(z)}-q_B\Big)\Big(q_A-\eta_{A}(z)\Big)+(1-q_A)(1-q_B)}{\Big(\frac{1}{\eta_{A}(z)}-q_A\Big)\Big(\frac{1}{\eta_{B}(z)}-q_B\Big)-(1-q_A)(1-q_B)}\;,
\end{equation}
\begin{equation}\label{eq38}
\Lambda_B(z)=\frac{(1-q_A)\Big(\frac{1}{\eta_{B}(z)}-\eta_{B}(z)\Big)}{\Big(\frac{1}{\eta_{A}(z)}-q_A\Big)\Big(\frac{1}{\eta_{B}(z)}-q_B\Big)-(1-q_A)(1-q_B)}\;,
\end{equation}
\begin{equation}\label{eq39}
\eta_{i}(z)=\frac{1+R_iz-\sqrt{(1+R_iz)^2-z^2}}{z}\;.
\end{equation}

As it was shown in the previous section, the discrete model leads to diffusion-absorption equations by the appropriate choice of the function $\hat{\omega}(s)$, more specifically by the function $v(s)$. For a membrane system, the choice of the functions $v_A(s)$ and $v_B(s)$ leads not only to the diffusion--absorption equations, but also to the boundary conditions at a thin membrane. 		
The Laplace transforms of the Green's functions for continuous time and discrete spatial variable are expressed by the formula
\begin{eqnarray}\label{eq40} 
  \hat{P}_i(m,s;m_0)=\frac{1-\hat{\omega}_i(s)}{s} S_i\left(m,\{\hat{\omega}_A(s),\hat{\omega}_B(s)\};m_0\right)\;, 
\end{eqnarray}
$i=A,B$, where $\hat{\omega}_i(s)=1/[1+\epsilon^2 v_i(s)/(2D_i)]$, 
the symbol $\{\hat{\omega}_A(s),\hat{\omega}_B(s)\}$ denotes that both functions $\hat{\omega}_A(s)$ and $\hat{\omega}_B(s)$ are involved into the functions $S_A$ and $S_B$ instead of the variable $z$ according to the following rule \cite{tk2,tk3}
\begin{equation}\label{eq42}  
    \eta_i(z)\rightarrow \eta_i(\hat{\omega}_i(s))\;,
\end{equation}
which also provides
\begin{equation}\label{eq43}
  \sqrt{(1+R_i z)^2-z^2} \rightarrow \sqrt{(1+R_i\hat{\omega}_i(s))^2-\hat{\omega}_i^2(s)}\;.
\end{equation}
The above rules were derived using the first passage time distribution $F(N,t;m_0)$ that the particle achieves the point $N$ first time at time $t$ starting from the point $m_0<N$. In terms of the Laplace transform this function reads 
\begin{eqnarray}\label{eq44}
\hat{F}(N,s;m_0)\equiv \frac{S_A(N,\{\hat{\omega}_A(s),\hat{\omega}_B(s)\};m_0)-\delta_{N,m_0}}{S_A(N,\{\hat{\omega}_A(s),\hat{\omega}_B(s)\};N)}\\
=\eta^{N-m_0}_A(\{\hat{\omega}_A(s),\hat{\omega}_B(s)\}).\nonumber
\end{eqnarray} 
Since this distribution is controlled by the function $\eta_A$ only and all steps are performed in the region $A$, the function $\eta_A$ should depend on the function $\hat{\omega}_A$ only when moving to continuous time. Similarly, the function $\eta_B$ depends on the function $\hat{\omega}_B$ only.

Moving to the continuous spatial variable we use the procedure presented in Sec. \ref{SecII} and Eqs. (\ref{eq35})--(\ref{eq43}). We get 
\begin{widetext}
\begin{eqnarray}\label{eq45}
\hat{P}_A(x,s;x_0)=\frac{v_A(s)}{2D_A s\sqrt{\kappa_A^2+\frac{v_A(s)}{D_A}}}\Big[{\rm e}^{-|x-x_0|\sqrt{\kappa_A^2+\frac{v_A(s)}{D_A}}}
+\Lambda_A(s){\rm e}^{-(2x_N-x-x_0)\sqrt{\kappa_A^2+\frac{v_A(s)}{D_A}}}\Big],
\end{eqnarray}
\begin{eqnarray}\label{eq46}
\hat{P}_B(x,s;x_0)=\frac{v_B(s)}{2D_B s\sqrt{\kappa_B^2+\frac{v_B(s)}{D_B}}}\;\Lambda_B(s)
\;{\rm e}^{-(x_N-x_0)\sqrt{\kappa_A^2+\frac{v_A(s)}{D_A}}}{\rm e}^{-(x-x_N)\sqrt{\kappa_B^2+\frac{v_B(s)}{D_B}}},
\end{eqnarray}
where $x_N=N\epsilon$,
\begin{equation}\label{eq47}
\Lambda_A(s)=\frac{(1-q_B)\sqrt{\kappa_A^2+\frac{v_A(s)}{D_A}}-(1-q_A)\sqrt{\kappa_B^2+\frac{v_B(s)}{D_B}}+\epsilon\sqrt{\kappa_A^2+\frac{v_A(s)}{D_A}}\sqrt{\kappa_B^2+\frac{v_B(s)}{D_B}}}{(1-q_B)\sqrt{\kappa_A^2+\frac{v_A(s)}{D_A}}+(1-q_A)\sqrt{\kappa_B^2+\frac{v_B(s)}{D_B}}+\epsilon\sqrt{\kappa_A^2+\frac{v_A(s)}{D_A}}\sqrt{\kappa_B^2+\frac{v_B(s)}{D_B}}}\;,
\end{equation}
\begin{equation}\label{eq48}
\Lambda_B(s)=\frac{2(1-q_A)\sqrt{\kappa_B^2+\frac{v_B(s)}{D_B}}}{(1-q_B)\sqrt{\kappa_A^2+\frac{v_A(s)}{D_A}}+(1-q_A)\sqrt{\kappa_B^2+\frac{v_B(s)}{D_B}}+\epsilon\sqrt{\kappa_A^2+\frac{v_A(s)}{D_A}}\sqrt{\kappa_B^2+\frac{v_B(s)}{D_B}}}\;.
\end{equation}
\end{widetext}
We note that $\Lambda_A(s)+\Lambda_B(s)=1$. To shorten the notation, in the following we will use the functions $\Lambda(s)\equiv \Lambda_B(s)$, then $\Lambda_A(s)=1-\Lambda(s)$.

The functions (\ref{eq45}) and (\ref{eq46}) fulfil the equation 
\begin{eqnarray}\label{eq49}
\frac{v_i(s)}{s}\left[s\hat{P}_i(x,s;x_0)-P_i(x,0;x_0)\right]\\=D_i\Bigg[\frac{\partial^2 \hat{P}_i(x,s;x_0)}{\partial x^2}
-\kappa_i^2\hat{P}_i(x,s;x_0)\Bigg],\nonumber
\end{eqnarray}
the Laplace transform of the flux $J_i$ is 
\begin{equation}\label{eq50}
\hat{J}_i(x,s;x_0)=-D_i\frac{s}{v_i(s)}\frac{\partial \hat{P}_i(x,s;x_0)}{\partial x},
\end{equation}
$i=A,B$.
Combining the values of functions $\hat{P}_A$, $\hat{P}_B$, $\hat{J}_A$, and $\hat{J}_B$ calculated for $x=x_N$ from Eqs. (\ref{eq45}), (\ref{eq46}), and (\ref{eq50}), we get the following boundary conditions at the thin membrane
\begin{equation}\label{eq51}
\hat{J}_A(x_N^-,s;x_0)=\hat{J}_B(x_N^+,s;x_0)\;,
\end{equation}
\begin{eqnarray}\label{eq52}
\frac{D_A}{v_A(s)}\sqrt{\kappa_A^2+\frac{v_A(s)}{D_A}}\hat{P}_A(x_N^-,s;x_0)\\ 
=\frac{D_B}{v_B(s)}\sqrt{\kappa_B^2+\frac{v_B(s)}{D_B}}\left(\frac{2-\Lambda(s)}{\Lambda(s)}\right)\hat{P}_B(x_N^+,s;x_0).\nonumber
\end{eqnarray}
The boundary conditions depend on the dimensional parameters. From Eq. (\ref{eq10}) it follows that the parameter $\mu$ is given in the units of $1/[v(s)]$, where $[v(s)]$ denotes the dimension of the function $v(s)$; for subdiffusion we have $[v(s)]=[s^\alpha]=1/(second)^\alpha$. Eq. (\ref{eq14}) shows that the unit of diffusion coefficient is $m^2[v(s)]$, and from Eq. (\ref{eq15}) we conclude that the unit of absorption coefficient $\kappa$ is $1/m$. Eq. (\ref{eq48}) shows that the function $\Lambda(s)$ is dimensionless. The above facts ensure that the dimensions of the right and left sides of the boundary condition Eq. (\ref{eq51}) are the same, the same applies to Eq. (\ref{eq52}).

Boundary condition Eq. (\ref{eq52}) can be applied in a system with continuous variables if we define the membrane permeability coefficients for a continuous system $\gamma_A$ and $\gamma_B$ and relate them to the probabilities $q_A$ and $q_B$.
We illustrate this problem considering diffusion in a system in which a thin symmetrical membrane separates two identical media. In this case there is assumed $q_A=q_B\equiv q$, $v_A(s)=v_B(s)\equiv v(s)$, $\kappa_A=\kappa_B\equiv \kappa$, and $D_A=D_B\equiv D$. Then, we get from Eq. (\ref{eq48})
\begin{equation}\label{eq53}
\Lambda(s)\equiv\Lambda_B(s)=\frac{2(1-q)}{2(1-q)+\epsilon \sqrt{\kappa^2+\frac{v(s)}{D}}}\;.
\end{equation}
If we assume that the probability of the particle's passing through a partially permeable membrane $1-q$, $0<q<1$, does not depend on the parameter $\epsilon$, then we have $\Lambda(s)\rightarrow 1$ in the limit of small $\epsilon$. In this case, the Green's functions Eqs. (\ref{eq45}) and (\ref{eq46}) take the form of the Green's function for a homogeneous system without a membrane Eq. (\ref{eq16}). It means that the membrane does not show its selective properties. The reason for this is as follows. The mean frequency of particle's jumps between neighbouring sites is $\nu(t)=d\left\langle n(t)\right\rangle/dt$, where $\left\langle n(t)\right\rangle$ is the number of steps over time interval $[0,t]$. We get
\begin{equation}\label{eq53a}
\nu(t)=\mathcal{L}^{-1}\Big[\frac{2D\hat{\omega}(s)}{\epsilon^2 v(s)}\Big].
\end{equation} 
Eq. (\ref{eq53a}) provides $\nu(t)\rightarrow\infty$ in the limit of small $\epsilon$. Then, the probability that a particle which tries to pass the partially permeable membrane `infinite times' in every finite time interval passes through the membrane is equal to one. To avoid such a non-physical situation, we use the following procedure when moving to a continuous spatial variable. The permeability properties of membrane are described by the function $\Lambda(s)$. This function should be independent of the parameter $\epsilon$. From Eq. (\ref{eq53}) it follows that this is possible only if $1-q\sim \epsilon$. Thus, in general the parameter $q$ can depend on $\epsilon$. Guided by the result presented above we suppose that
\begin{equation}\label{eq54}
1-q_A=\frac{\epsilon^{\sigma_A}}{\gamma_A}\;,\;1-q_B=\frac{\epsilon^{\sigma_B}}{\gamma_B},
\end{equation}
where $\sigma_A$ and $\sigma_B$ are parameters as yet to be determined, $\gamma_A$ and $\gamma_B$ are the membrane permeability coefficients defined for the system with continuous variables. Since $0\leq q_{A,B}\leq 1$, we have $\sigma_{A,B}\geq 0$ and $\gamma_{A,B}>0$. 

For a one-sided fully permeable membrane the Green's functions and boundary conditions are given by Eqs. (\ref{eq45})--(\ref{eq48}) with $q_A=0$ or $q_B=0$. When the boundary between media does not make any obstacle for the diffusing particles we have $q_A=q_B=0$. Below we will consider the above mentioned cases separately. The Green's functions are still given by Eqs. (\ref{eq45}) and (\ref{eq46}), but the function $\Lambda(s)$ is different for these cases.

\subsection{The case of $q_A\neq 0$ and $q_B\neq 0$\label{SecIIIA}}

For $1<\sigma_A$ and $1<\sigma_B$ we get $\Lambda=0$, so we obtain the Green's function for the system with fully reflecting wall. For $\sigma_A<1$ and $\sigma_B<1$ the membrane `vanishes' in the case of symmetrical system. The cases of $\sigma_A>1$, $\sigma_B<1$ and $\sigma_A<1$, $\sigma_B>1$ also provides non-physical results $\Lambda(s)\equiv -1$ and $\Lambda(s)\equiv 1$, respectively. Thus, we get $\sigma_A=\sigma_B=1$ and
\begin{equation}\label{eq55}
1-q_A=\frac{\epsilon}{\gamma_A}\;,\; 1-q_B=\frac{\epsilon}{\gamma_B}.
\end{equation}
Taking into account Eqs. (\ref{eq48}) and (\ref{eq55}) we obtain
\begin{widetext}
\begin{equation}\label{eq56}
\Lambda(s)=\frac{2\gamma_B\sqrt{\kappa_B^2+\frac{v_B(s)}{D_B}}}{\gamma_A\sqrt{\kappa_A^2+\frac{v_A(s)}{D_A}}+\gamma_B\sqrt{\kappa_B^2+\frac{v_B(s)}{D_B}}+\gamma_A\gamma_B\sqrt{\kappa_A^2+\frac{v_A(s)}{D_A}}\sqrt{\kappa_B^2+\frac{v_B(s)}{D_B}}}\;.
\end{equation}
\end{widetext}

\subsection{The case of $q_A= 0$ and $q_B\neq 0$\label{SecIIIB}}

For $\sigma_B>0$ we get $\Lambda(s)=2$ when $\epsilon\rightarrow 0$ and we obtain the Green's function for the system with fully absorbing wall. Thus, we assume $\sigma_B=0$, so we get
\begin{equation}\label{eq57a} 
1-q_B=\frac{1}{\gamma_B}
\end{equation} 
and
\begin{equation}\label{eq57}
\Lambda(s)=\frac{2\gamma_B\sqrt{\kappa_B^2+\frac{v_B(s)}{D_B}}}{\sqrt{\kappa_A^2+\frac{v_A(s)}{D_A}}+\gamma_B\sqrt{\kappa_B^2+\frac{v_B(s)}{D_B}}}\;.
\end{equation}

\subsection{The case of $q_A\neq 0$ and $q_B= 0$\label{SecIIIC}}

For $\sigma_A>0$ we get $\Lambda(s)=0$, which provides the Green's function for the system with fully reflecting membrane. Thus, we suppose $\sigma_A=0$ which provides 
\begin{equation}\label{eq58a}
1-q_A=\frac{1}{\gamma_A}
\end{equation} 
and 
\begin{equation}\label{eq58}
\Lambda(s)=\frac{2\sqrt{\kappa_B^2+\frac{v_B(s)}{D_B}}}{\gamma_A\sqrt{\kappa_A^2+\frac{v_A(s)}{D_A}}+\sqrt{\kappa_B^2+\frac{v_B(s)}{D_B}}}\;.
\end{equation}

\subsection{The case of $q_A= 0$ and $q_B= 0$\label{SecIIID}}

In this case we have
\begin{equation}\label{eq59}
\Lambda(s)=\frac{2\sqrt{\kappa_B^2+\frac{v_B(s)}{D_B}}}{\sqrt{\kappa_A^2+\frac{v_A(s)}{D_A}}+\sqrt{\kappa_B^2+\frac{v_B(s)}{D_B}}}\;.
\end{equation}
This case is considered in \cite{tk3} for the `classical' subdiffusion with absorption process in both media $A$ and $B$.

\section{How to solve the system of diffusion equations for any initial condition\label{SecIV}}

The above considerations were performed assuming that $x_0<x_N$. Till now the Green's functions have been marked with one index indicating to which region the point $x$ belongs. 
However, when the membrane is asymmetrical the boundary condition can depend on which side of the membrane a particle starts its motion, see the discussion in \cite{tk}. An example is a thin membrane that is fully impenetrable to particles moving from the region $A$ to the region $B$ and partially permeable when particles move in the opposite direction. Then, for particles initially located in $A$ the boundary condition at the membrane is just as for fully reflecting wall whereas for particles starting form the region $B$ the boundary condition is as for partially absorbing wall. 

We consider the situation in which the initial position of the particle may be in both regions $A$ and $B$. Then, the Green's functions are labelled by two indexes $ij$ which show the regions to which $x$ and $x_0$ points belong, respectively. 
In the new notation the functions $\hat{P}_{AA}(x,s;x_0)$ and $\hat{P}_{BA}(x,s;x_0)$ are expressed by Eqs. (\ref{eq45}) and (\ref{eq46}), respectively. Due to the symmetry arguments, assuming $x_0>x_N$ we can obtain the Green's functions and boundary conditions making the following changes $(x-x_0,x-x_N,x_0-x_N)\leftrightarrow(x_0-x,x_N-x,x_N-x_0)$ and $A\leftrightarrow B$ of all indexes occurring in Eqs. (\ref{eq45})--(\ref{eq48}); we note that the change $\Lambda_A(s)\leftrightarrow\Lambda_B(s)$ is equivalent to $1-\Lambda(s)\leftrightarrow\Lambda(s)$. 

In the following we assume that particles move independently of one another and that the membrane permeability parameters do not depend on the concentration of the diffusing particles. We denote
\begin{eqnarray}\label{eq60}
  C(x,t)&=&\left\{\begin{array}{ll}
      C_{A}(x,t)\;,&x<x_N\;,\\
      C_{B}(x,t)\;,&x>x_N\;.
    \end{array}\right.
\end{eqnarray}
We assume the initial condition as follows
\begin{eqnarray}\label{eq61}
  C(x,0)&=&\left\{\begin{array}{ll}
      C_{0A}(x)\;,&x<x_N\;,\\
      C_{0B}(x)\;,&x>x_N\;.
    \end{array}\right.
\end{eqnarray}

Let $C_{AA}(x,t)$ and $C_{BA}(x,t)$ denote the solutions defined in the regions $A$ and $B$, respectively, generated by the particles located initially in the region $A$, i.e. by the following initial condition
\begin{eqnarray}\label{eq62}
  \left\{\begin{array}{ll}
      C_{AA}(x,0)=C_{0A}(x)\;,&x<x_N\;,\\
      C_{BA}(x,0)=0\;,&x>x_N\;.
    \end{array}\right.
\end{eqnarray}
Since it is assumed that the particles move independently of one another, the concentrations can be calculated by means of the formula
\begin{equation}\label{eq63}
C_{iA}(x,t)=\int_{-\infty}^{x_N} P_{iA}(x,t;x_0)C_A(x_0)dx_0.
\end{equation}
The functions $C_{AB}(x,t)$ and $C_{BB}(x,t)$, generated by the initial condition
\begin{eqnarray}\label{eq64}
  \left\{\begin{array}{ll}
      C_{AB}(x,0)=0\;,&x<x_N\;,\\
      C_{BB}(x,0)=C_{0B}(x)\;,&x>x_N\;,
    \end{array}\right.
\end{eqnarray}
can be calculated as follows
\begin{equation}\label{eq65}
C_{iB}(x,t)=\int_{x_N}^\infty P_{iB}(x,t;x_0)C_{0B}(x_0)dx_0,
\end{equation}
$i=A,B$.
From Eqs. (\ref{eq23}), (\ref{eq63}), and (\ref{eq65}) we obtain
\begin{eqnarray}\label{eq67}
\int_0^t F_i(t-t')\frac{\partial C_{ij}(x,t')}{\partial t'}dt'=D_i\Bigg[\frac{\partial^2 C_{ij}(x,t)}{\partial x^2}\\
-\kappa_i^2 C_{ij}(x,t)\Bigg],\nonumber
\end{eqnarray}
where $F_i(t)=\mathcal{L}^{-1}[v_i(s)/s]$, $i,j\in\{A,B\}$.
The solutions to the considered equations are superposition of the partial solutions described above
\begin{eqnarray}\label{eq66}
  \left\{\begin{array}{ll}
      C_{A}(x,t)=C_{AA}(x,t)+C_{AB}(x,t)\;,&x<x_N\;,\\
      C_{B}(x,t)=C_{BA}(x,t)+C_{BB}(x,t)\;,&x>x_N\;.
    \end{array}\right.
\end{eqnarray}

Due to the complex form of the equations and boundary conditions in the time domain, it is convenient to find the solutions in terms of the Laplace transform. From Eq. (\ref{eq67}) we get 
\begin{eqnarray}\label{eq68}
\frac{v_i(s)}{s}\left[s\hat{C}_{ij}(x,s)-C_{ij}(x,0)\right]\\=D_i\Bigg[\frac{\partial^2 \hat{C}_{ij}(x,s)}{\partial x^2}
-\kappa_i^2\hat{C}_{ij}(x,s)\Bigg].\nonumber
\end{eqnarray}
The Laplace transform of the flux $J_{ij}(x,t)$ reads 
\begin{equation}\label{eq69}
\hat{J}_{ij}(x,s)=-D_i\frac{s}{v_i(s)}\frac{\partial \hat{C}_{ij}(x,s)}{\partial x},
\end{equation}
$i,j\in \{A,B\}$.
From Eqs. (\ref{eq51}), (\ref{eq52}), (\ref{eq63}), and (\ref{eq65}) we get the boundary conditions at the thin membrane for the functions $\hat{C}_{AA}$ and $\hat{C}_{BA}$
\begin{equation}\label{eq70}
\hat{J}_{AA}(x^-_N,s)=\hat{J}_{BA}(x^+_N,s)\;,
\end{equation}
\begin{eqnarray}\label{eq71}
\frac{D_A}{v_A(s)}\sqrt{\kappa_A^2+\frac{v_A(s)}{D_A}}\hat{C}_{AA}(x^-_N,s)\\ 
=\frac{D_B}{v_B(s)}\sqrt{\kappa_B^2+\frac{v_B(s)}{D_B}}\left(\frac{2-\Lambda(s)}{\Lambda(s)}\right)\hat{C}_{BA}(x^+_N,s).\nonumber
\end{eqnarray}
Using the symmetry rule, we obtain the boundary condition at the thin membrane for the functions $\hat{C}_{AB}$ and $\hat{C}_{BB}$
\begin{equation}\label{eq72}
\hat{J}_{AB}(x^-_N,s)=\hat{J}_{BB}(x^+_N,s)\;,
\end{equation}
\begin{eqnarray}\label{eq73}
\frac{D_A}{v_A(s)}\sqrt{\kappa_A^2+\frac{v_A(s)}{D_A}}\left(\frac{1+\Lambda(s)}{1-\Lambda(s)}\right)\hat{C}_{AB}(x^-_N,s)\\
=\frac{D_B}{v_B(s)}\sqrt{\kappa_B^2+\frac{v_B(s)}{D_B}}\hat{C}_{BB}(x^+_N,s).\nonumber 
\end{eqnarray}

In summary, the method of solving the system of diffusion--absorption equations for a system with a thin membrane is as follows
\begin{enumerate}
	\item Find the solutions $\hat{C}_{AA}$ and $\hat{C}_{BA}$ to Eq. (\ref{eq68}) for $i=A$ with the initial condition Eq. (\ref{eq62}) and the boundary conditions at the membrane Eqs. (\ref{eq70}) and (\ref{eq71}).
	\item Find the solutions $\hat{C}_{AB}$ and $\hat{C}_{BB}$ to Eq. (\ref{eq68}) for $i=B$ with the initial condition Eq. (\ref{eq64}) and the boundary conditions at the membrane Eqs. (\ref{eq72}) and (\ref{eq73}).
	\item Find the functions $\hat{C}_A$ and $\hat{C}_B$ using the Laplace transform of Eq. (\ref{eq66}).
	\item To obtain the final solutions calculate the inverse Laplace transform of $\hat{C}_A$ and $\hat{C}_B$. 
\end{enumerate}

Two boundary conditions should be additionally given at points distant from the membrane, for example, one assumes finite solutions when $x\rightarrow\pm\infty$ or zero values of particles' fluxes at the external walls of the vessel. The inverse Laplace transforms can be calculated using standard formulas supplemented with formulas presented in the Appendix. If it is not possible to accurately calculate the inverse Laplace transform, the approximation of small parameter $s$ can be used; this approximation corresponds to the long time limit in the time domain.

\section{Absorption of diffusing substance by subdiffusive medium\label{SecV}}

As an example, we consider the diffusion of particles from a medium $A$ in which subdiffusion or normal diffusion occurs to a porous medium $B$ where absorption can be present. Let us also assume that the particle which try to pass the border between media moving form the medium $B$ to $A$ can do it without any obstacle, but when the particle moves in the opposite direction it can pass the border with some probability. The particle can be stopped at the border in the latter case if they do not go directly to one of the channels located in the porous medium. Thus, we assume that $\kappa_A=0$, $q_A\neq 0$, and $q_B=0$. 
The example of this process is the water purification process by halloysite or kaolinite (medium $B$) submerged in water (medium $A$) \cite{lz,zavl}. The impurities diffuse inside the clay medium, in which adsorption of diffusing particles can occur with a certain probability. 

We assume that $x_0<x_N$ at the initial moment.
Assuming that 
\begin{equation}\label{eq74}
\gamma_A\sqrt{\frac{v_A(s)}{D_A}}\ll \sqrt{\kappa_B^2+\frac{v_B(s)}{D_B}},
\end{equation}
the function $\Lambda(s)$ can be approximated as
\begin{equation}\label{eq75}
\Lambda(s)=2\left[1-\frac{\gamma_A\sqrt{\frac{v_A(s)}{D_A}}}{\sqrt{\kappa_B^2+\frac{v_B(s)}{D_B}}}\right].
\end{equation}
Then the Laplace transforms of the Green's functions read
\begin{widetext}
\begin{eqnarray}\label{eq76}
\hat{P}_{AA}(x,s;x_0)=\frac{\sqrt{v_A(s)}}{2\sqrt{D_A} s}\Bigg[{\rm e}^{-|x-x_0|\sqrt{\frac{v_A(s)}{D_A}}}
-{\rm e}^{-(2x_N-x-x_0)\sqrt{\frac{v_A(s)}{D_A}}}\Bigg]+\frac{\gamma_A v_A(s)}{D_A s\sqrt{\kappa_B^2 +\frac{v_B(s)}{D_B}}}\;{\rm e}^{-(2x_N-x-x_0)\sqrt{\frac{v_A(s)}{D_A}}},
\end{eqnarray}
\begin{eqnarray}\label{eq77}
\hat{P}_{BA}(x,s;x_0)=\frac{v_B(s)}{D_B s\sqrt{\kappa_B^2+\frac{v_B(s)}{D_B}}}\left(1-\frac{\gamma_A \sqrt{v_A(s)}}{\sqrt{D_A\left(\kappa_B^2+\frac{v_B(s)}{D_B}\right)}}\right)\;{\rm e}^{-(x_N-x_0)\sqrt{\frac{v_A(s)}{D_A}}}\;{\rm e}^{-(x-x_N)\sqrt{\kappa_B^2+\frac{v_B(s)}{D_B}}}.
\end{eqnarray}
\end{widetext}
We suppose that at the initial moment there is homogeneous solution in the medium $A$ whereas the medium $B$ is free of the diffusing substance, the initial condition is
\begin{eqnarray}\label{eq78}
  \left\{\begin{array}{ll}
      C_{AA}(x,0)=C_{0}\;,&x<x_N\;,\\
      C_{BA}(x,0)=0\;,&x>x_N\;.
    \end{array}\right.
\end{eqnarray}
The concentrations $\hat{C}_{AA}$ and $\hat{C}_{BA}$ can be calculated using Eqs. (\ref{eq63}) and (\ref{eq76})--(\ref{eq78}). The temporal evolution of the amount of substance which leave the region $A$ is $W_A(t,s)=\int_{-\infty}^{x_N}[C_{AA}(x,0)-C_{AA}(x,t)]dx$ and the amount of substance which is in the part $B$ reads $W_B(t)=\int_{x_N}^\infty C_{BA}(x,t)dx$. The amount of substance absorbed in the time interval $[0,t]$ is $W(t)=W_A(t)-W_B(t)$.
In terms of the Laplace transform we get 
\begin{equation}\label{eq79}
\hat{W}_A(s)=\frac{C_0}{s}\left[\frac{\sqrt{D_A}}{\sqrt{v_A(s)}}-\frac{\gamma_A}{\sqrt{\kappa_B^2+\frac{v_B(s)}{D_B}}}\right],
\end{equation}
\begin{equation}\label{eq80}
\hat{W}_B(s)=\frac{C_0 v_B(s)}{sD_B\left(\kappa_B^2+\frac{v_B(s)}{D_B}\right)}\left[\frac{\sqrt{D_A}}{\sqrt{v_A(s)}}-\frac{\gamma_A}{\sqrt{\kappa_B^2+\frac{v_B(s)}{D_B}}}\right].
\end{equation}
Below we will derive the Green's functions and the functions $W_A(t)$, $W_B(t)$, and $W(t)$ in the long time limit. We will consider the cases of subdiffusion and slow subdiffusion in the medium $B$. For both cases we assume $v_A(s)=s^{\alpha_A}$.

\subsection{Subdiffusion in the medium $B$\label{SecVA}}

We assume that $v_B(s)=s^{\alpha_B}$ and $\alpha_A>\alpha_B$. To calculate $P_B$ we use the series ${\rm e}^{-(x_N-x_0)\sqrt{s^\alpha_A/D_A}}=\sum_{k=0}^\infty \frac{1}{k!}\left(\frac{-(x_N-x_0)s^{\alpha_A}}{\sqrt{D_A}}\right)^k$. Instead of Eq. (\ref{eq74}), we use here the stronger condition $\gamma_A\sqrt{\frac{s^{\alpha_A}}{D_A}}\ll \kappa_B$, which, according to the relation (\ref{eqa4}) from the Appendix I, gives $t\gg(\kappa_B^2 D_A/\gamma_A)^{1/\alpha_A}$. Under this condition we get 
\begin{widetext}
\begin{eqnarray}\label{eq81}
P_{AA}(x,t;x_0)=\frac{1}{2\sqrt{D_A}}\Bigg[f_{\alpha_A/2-1,\alpha_A/2}\left(t;\frac{|x-x_0|}{\sqrt{D_A}}\right)-f_{\alpha_A/2-1,\alpha_A/2}\left(t;\frac{2x_N-x-x_0}{\sqrt{D_A}}\right)\Bigg]\\
+\frac{\gamma_A}{D_A\kappa_B}\Bigg[f_{\alpha_A-1,\alpha_A/2}\left(t;\frac{2x_N-x-x_0}{\sqrt{D_A}}\right)-\frac{1}{2\kappa_B^2 D_B}f_{\alpha_A+\alpha_B-1,\alpha_A/2}\left(t;\frac{2x_N-x-x_0}{\sqrt{D_A}}\right)\Bigg],\nonumber
\end{eqnarray}
\begin{eqnarray}\label{eq82}
P_{BA}(x,t;x_0)=\frac{{\rm e}^{-\kappa_B(x-x_N)}}{D_B\kappa_B}\sum_{n=0}^\infty\frac{(x_0-x_N)^n}{n!(\sqrt{D_A})^n}\Bigg[f_{\alpha_B-1+n\alpha_A/2,\alpha_B}\left(t;\frac{x-x_N}{2\kappa_B D_B}\right)-\frac{1}{2\kappa_B^2 D_B}f_{2\alpha_B-1+n\alpha_A/2,\alpha_B}\left(t;\frac{x-x_N}{2\kappa_B D_B}\right)\\
-\frac{\gamma_A}{\sqrt{D_A}\kappa_B}f_{\alpha_A/2+\alpha_B-1+n\alpha_A/2,\alpha_B}\left(t;\frac{x-x_N}{2\kappa_B D_B}\right)
+\frac{\gamma^2_A}{D_A \kappa^2_B}f_{\alpha_A+\alpha_B-1+n\alpha_A/2,\alpha_B}\left(t;\frac{x-x_N}{2\kappa_B D_B}\right)\Bigg],\nonumber
\end{eqnarray}
\end{widetext}
where the functions $f_{\alpha,\nu}$ are expressed by Eq. (\ref{eqa1}) from the Appendix I, and
\begin{eqnarray}\label{eq83}
W_A(t)=C_0\Bigg[\frac{\sqrt{D_A}\;t^{\alpha_A/2}}{\Gamma(1+\alpha_A/2)}-\frac{\gamma_A}{\kappa_B}+\frac{\gamma_A}{2D_B\kappa_B^3t^{\alpha_B}}\Bigg],
\end{eqnarray}
\begin{eqnarray}\label{eq84}
W_B(t)=\frac{C_0}{D_B\kappa_B^3 t^{\alpha_B}}\Bigg[\frac{\sqrt{D_A}\;t^{\alpha_A/2}}{\Gamma(1+\alpha_A/2-\alpha_B)}\\-\frac{\gamma_A}{\kappa_B\Gamma(1-\alpha_B)}\Bigg].\nonumber
\end{eqnarray}
From Eqs. (\ref{eq83}) and (\ref{eq84}) we get
\begin{eqnarray}\label{eq85}
W(t)=C_0\Bigg[\frac{\sqrt{D_A}\;t^{\alpha_A/2}}{\Gamma(1+\alpha_A/2)}-\frac{\gamma_A}{\kappa_B}\\
+\frac{\sqrt{D_A}t^{\alpha_A/2-\alpha_B}}{D_B\kappa_B^3\Gamma(1+\alpha_A/2-\alpha_B)}\Bigg].\nonumber
\end{eqnarray}

		\begin{figure}[h]
        \includegraphics[scale=0.4]{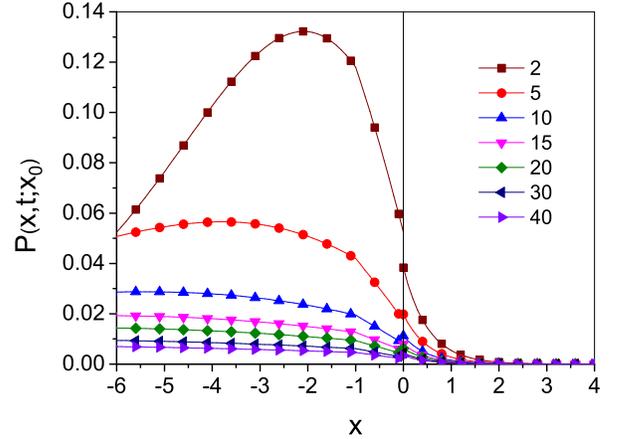}
				\caption{The plots of Green's functions Eqs. (\ref{eq81}) and (\ref{eq82}) for subdiffusion in the region $B$ calculated for $\alpha_A=0.9$, $\alpha_B=0.8$, $\kappa_B=2.0$ and for times given in the legend, all quantities are given in arbitrarily chosen units. \label{Fig3}}
      \end{figure}	
		\begin{figure}[h]
        \includegraphics[scale=0.4]{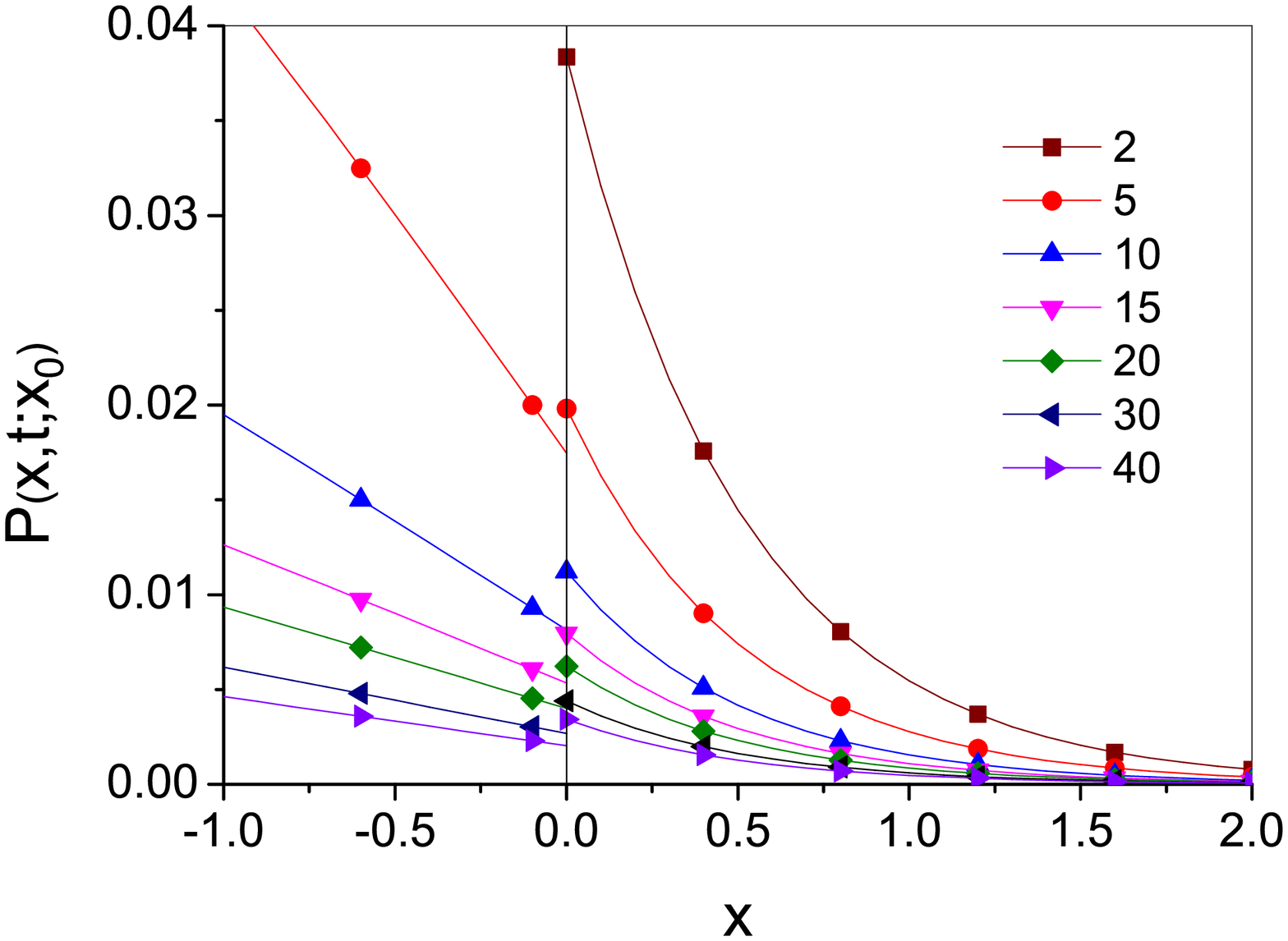}
				\caption{The fragment of the plot Fig. \ref{Fig3} made in the other scale.\label{Fig4}}
      \end{figure}
		\begin{figure}[h]
        \includegraphics[scale=0.4]{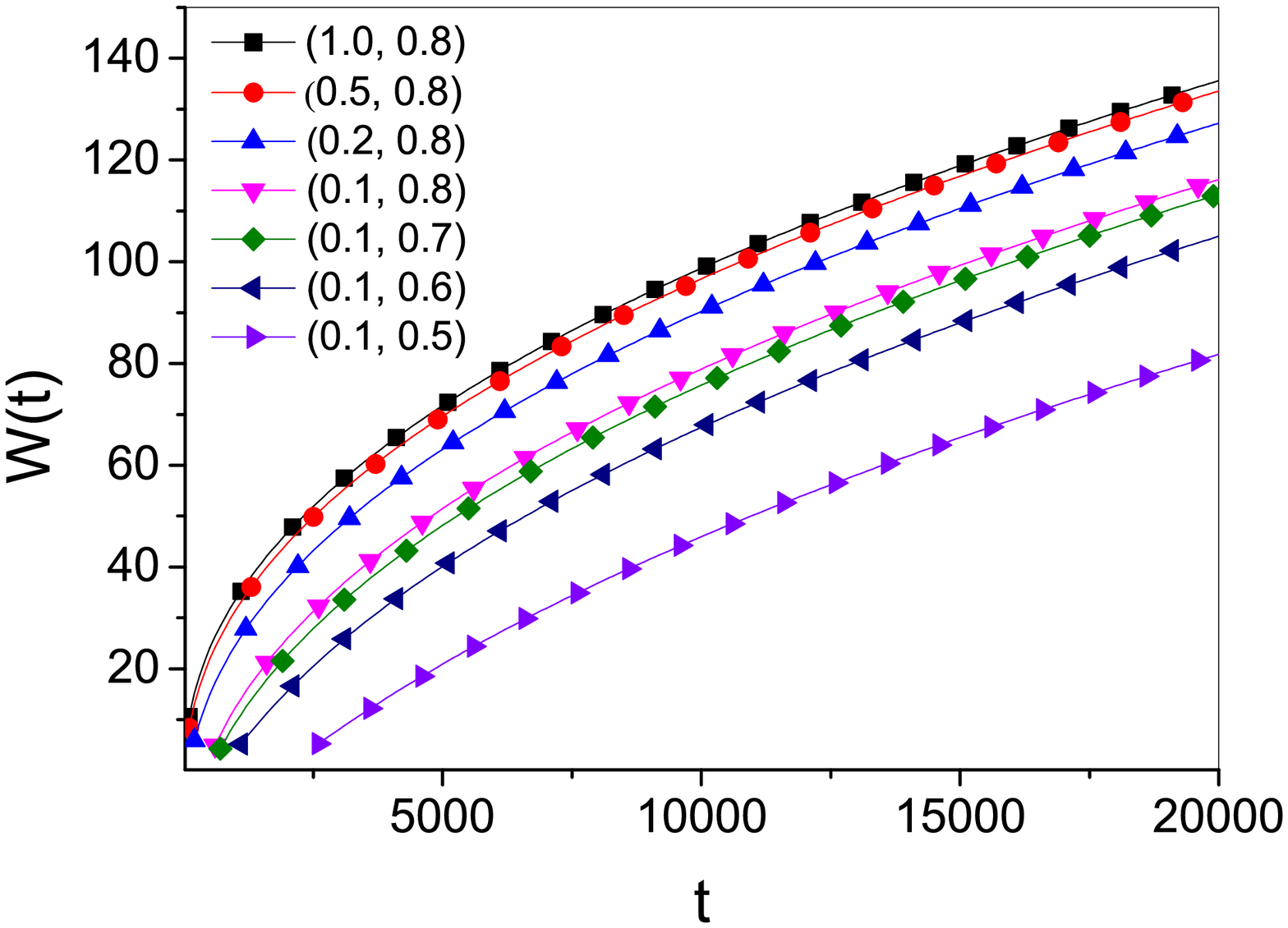}
				\caption{The plots of the function $W$ Eq. (\ref{eq85}) for subdiffusion in the medium $B$ for $\alpha_A=0.9$ and various $(\kappa_B,\alpha_B)$ given in the legend.\label{Fig5}}
      \end{figure}

\subsection{Slow subdiffusion in the medium $B$\label{SecVB}}

We assume $v_B(s)=1/{\rm ln}^r(1/s)$, $r>0$. Using Eqs. (\ref{eqa6}) and (\ref{eqa7}) from the Appendix I, we get in the limit of long time
\begin{eqnarray}\label{eq86}
P_{AA}(x,t;x_0)=\frac{1}{2\sqrt{D_A}}\Bigg[f_{\alpha_A/2-1,\alpha_A/2}\left(t;\frac{|x-x_0|}{\sqrt{D_A}}\right)\\
-f_{\alpha_A/2-1,\alpha_A/2}\left(t;\frac{2x_N-x-x_0}{\sqrt{D_A}}\right)\Bigg]\nonumber\\
+\frac{\gamma_A}{D_A\kappa_B\sqrt{\kappa_B^2+\frac{v_B(1/t)}{D_B}}}f_{\alpha_A-1,\alpha_A/2}\left(t;\frac{2x_N-x-x_0}{\sqrt{D_A}}\right),\nonumber
\end{eqnarray}
\begin{eqnarray}\label{eq87}
P_{BA}(x,t;x_0)=\frac{v_B(1/t)\;{\rm e}^{-(x-x_N)\sqrt{\kappa_B^2+\frac{v_B(1/t)}{D_B}}}}{D_B \sqrt{\kappa_B^2+\frac{v_B(1/t)}{D_B}}}\\
\times\Bigg[f_{-1,\alpha_A/2}\left(t;\frac{x_N-x_0}{\sqrt{D_A}}\right)\nonumber\\
-\frac{\gamma_A}{\sqrt{\kappa_B^2+\frac{v_B(1/t)}{D_B}}}f_{\alpha_A/2-1,\alpha_A/2}\left(t;\frac{x_N-x_0}{\sqrt{D_A}}\right)\Bigg],\nonumber
\end{eqnarray}
and
\begin{equation}\label{eq88}
W_A(t)=C_0\left[\frac{\sqrt{D_A}t^{\alpha_A/2}}{\Gamma(1+\alpha_A/2)}-\frac{\gamma_A}{\sqrt{\kappa_B^2+\frac{v_B(1/t)}{D_B}}}\right],
\end{equation}
\begin{eqnarray}\label{eq89}
W_B(t)=\frac{C_0 v_B(1/t)}{D_B\left(\kappa_B^2+\frac{v_B(1/t)}{D_B}\right)}\Bigg[\frac{\sqrt{D_A}t^{\alpha_A/2}}{\Gamma(1+\alpha_A/2)}\\
-\frac{\gamma_A}{\sqrt{\kappa_B^2+\frac{v_B(1/t)}{D_B}}}\Bigg].\nonumber
\end{eqnarray}
			\begin{figure}[h]
        \includegraphics[scale=0.4]{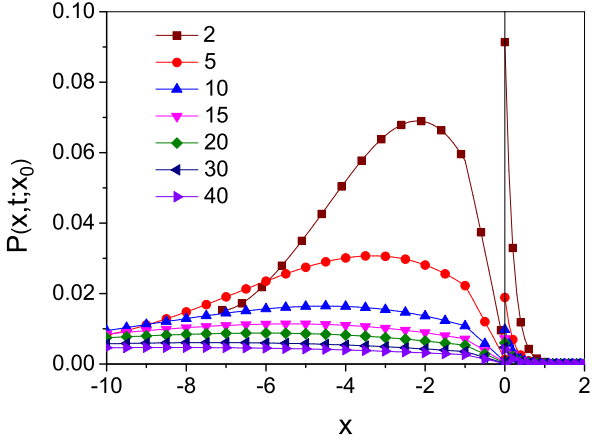}
				\caption{The plots of Green's functions Eqs. (\ref{eq86}) and (\ref{eq87}) for slow subdiffusion in the region $B$ calculated for $\alpha_A=0.9$, $r=2.0$, $\kappa_B=2.0$ and for times given in the legend.\label{Fig6}}
      \end{figure}
		\begin{figure}[h]
        \includegraphics[scale=0.4]{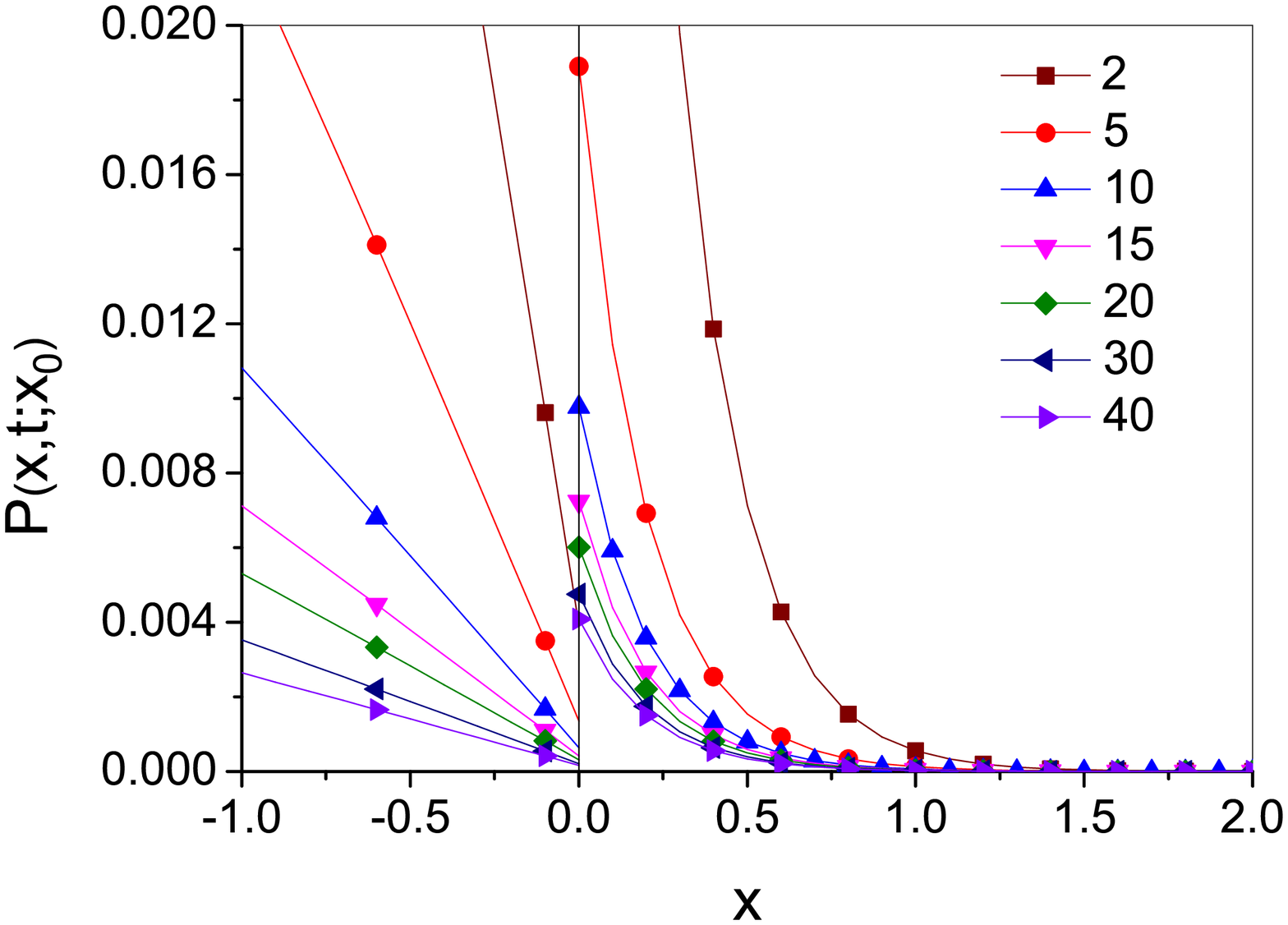}
				\caption{The fragment of the plot Fig. \ref{Fig6} made in the other scale.\label{Fig7}}
      \end{figure}		
		\begin{figure}[h]
        \includegraphics[scale=0.4]{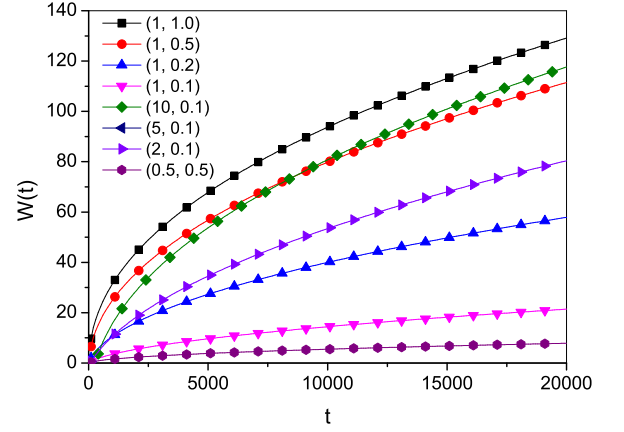}
				\caption{The plots of the function $W$ Eq. (\ref{eq90}) for slow subdiffusion in the medium $B$ for $\alpha_A=0.9$ and various $(r,\kappa_B)$ given in the legend.\label{Fig8}}
     \end{figure}
From Eqs. (\ref{eq88}) and (\ref{eq89}) we obtain 
\begin{eqnarray}\label{eq90}
W(t)=C_0\left[\frac{\sqrt{D_A}\;t^{\alpha_A/2}}{\Gamma(1+\alpha_A/2)}\Bigg(1-\frac{v_B(1/t)}{D_B\kappa_B^2+v_B(1/t)}\Bigg)\right.\\
\left.-\frac{\gamma_A}{\sqrt{\kappa_B^2+\frac{v_B(1/t)}{D_B}}}\right].\nonumber
\end{eqnarray}

The plots of Green's functions are presented in Figs. \ref{Fig3} and \ref{Fig4} for subdiffusion in the region $B$ and in Figs. \ref{Fig6} and \ref{Fig7} for slow subdiffusion in $B$; Figs. \ref{Fig4} and \ref{Fig7} are the fragments of Figs. \ref{Fig3} and \ref{Fig6}, respectively, made on a different scale. In Figs. \ref{Fig5} and \ref{Fig8} the plots of function $W$ for subdiffusion and slow subdiffusion are shown.  All plots are made for $x_N=0$, $x_0=-1.0$, $D_A=D_B=2.0$ and $\gamma_A=2.0$, values of the other parameters are given in the plot captions, all quantities are given in arbitrarily chosen units. The limit of long time is $t\gg 100$.

We note that $P_A(x,t;x_0)\rightarrow 0$ when $t\rightarrow\infty$. Thus, in the long time limit the medium $B$ influences the diffusion of particles in the medium $A$ in a similar way as the absorbing wall placed at $x_N$. The reason is that the mobility of particles in the medium $B$ is much smaller compared to the medium $A$, the effect is enhanced by absorption in the medium $B$. From the equations presented in this section we note that the functions $W$ and $W_A$ have the same asymptotic forms for long time for subdiffusion as well as for slow subdiffusion in the medium $B$. Analysing qualitatively the plots presented in Figs. \ref{Fig5} and \ref{Fig8} we deduce that the loss of substance in the medium $A$ is determined mainly by the absorption process in the medium $B$, and to a lesser extent by the mobility of particles in this medium.

\section{Final remarks\label{SecVI}}

The most important results presented in this paper are:
\begin{itemize}
	\item the general form of the Green's functions Eqs. (\ref{eq45}) and (\ref{eq46}) for a system consisting of two media separated by a thin membrane treated as a partially permeable wall; different kinds of diffusion--absorption processes can occur in the media,
	\item boundary conditions at the border between media, Eqs. (\ref{eq51}) and (\ref{eq52}),
	\item method of solving the diffusion equations for any initial conditions, presented in Sec. \ref{SecIV}.
\end{itemize}
The method provides the solution in terms of the Laplace transform. Usually, it is difficult to calculate the inverse Laplace transform of the obtained solutions. However, it is often possible to find the inverse Laplace transforms over the long time limit, which corresponds to the limit of small parameter $s$. Some useful formulas for calculating inverse Laplace transforms are shown in the Appendix.

Some specific remarks are as follows.
\begin{enumerate}
	\item The method presented in Sec. \ref{SecIV} can be generalized to model of diffusion with absorption in a multilayer system. In each layer, diffusion--absorption processes is described by Eqs. (\ref{eq68}) with the function $v_i$ which defined the kind of diffusion, the boundary conditions are expressed by Eqs. (\ref{eq70})--(\ref{eq73}).
	\item The boundary conditions Eqs. (\ref{eq70})--(\ref{eq73}) can be used to solve diffusion--absorption equations in two- or three dimensional space where the boundary conditions should be set in the direction normal to the surface separating different media.
	\item In the presented model we consider a thin membrane placed between the media. In practice, the membrane represents any obstacle that can, with certain probability, stop the diffusing particle. An example of this may be the transient layer formed at the gel-water or biofilm-water interface, see the discussion in \cite{tk2}.
	\item The interpretation of membrane permeability coefficients is based on the equation (\ref{eq54}), $\epsilon$ can be interpreted as the thickness of a thin membrane, $q$ can be calculated from a phenomenological model. In the simplest model there is $q=\Pi_P/\Pi$, where $\Pi_P$ is the area of the pores observed on the membrane surface and $\Pi$ is the total area of the membrane surface.
	\item The boundary conditions at a thin membrane for the system which consists of two diffusive media without absorption, derived from Eqs. (\ref{eq70}) and (\ref{eq71}) putting $v_A=v_B=s$, $D_A=D_B\equiv D$, and $\kappa_A=\kappa_B=0$, are the same as the boundary conditions obtained from experimental data \cite{kwl}. In this case the boundary condition (\ref{eq71}) reads in the time domain 
	\begin{eqnarray}\label{eq101}
	C_{AA}(x_N^-,t)=\Bigg(\frac{\gamma_A}{\gamma_B}+\frac{\gamma_A}{\sqrt{D}}\frac{\partial^{1/2}}{\partial t^{1/2}}\Bigg)C_{BA}(x_N^+,t).
	\end{eqnarray}
Although the normal diffusion process was considered, this boundary condition contains the Riemann--Liouville fractional time derivative of the order $1/2$.
	
	\item A frequently used boundary condition in a system without absorption in which a thin membrane is placed in a homogeneous medium, $v_A(s)=v_B(s)\equiv v(s)$ and $D_A=D_B\equiv D$, is 
	\begin{equation}\label{eq102}
	J_{B}(x_N^+,t;x_0)=\lambda_1 P_{A}(x_N^-,t;x_0)-\lambda_2 P_{B}(x_N^+,t;x_0),
	\end{equation}
the parameters $\lambda_1$ and $\lambda_2$ control the membrane permeability.  Assuming that the diffusive flux is continuous, the fundamental solutions to Eq. (\ref{eq49}) with $\kappa_A=\kappa_B=0$ for the boundary conditions Eqs. (\ref{eq51}) and the Laplace transform of Eq. (\ref{eq102}) are
\begin{eqnarray}\label{eq103}
\hat{P}_A(x,s;x_0)=\frac{1}{2s}\sqrt{\frac{v(s)}{D}}\Bigg[{\rm e}^{-|x-x_0|\sqrt{\frac{v(s)}{D}}}\\
+(1-\Xi(s)){\rm e}^{-(2x_N-x-x_0)\sqrt{\frac{v(s)}{D}}}\Bigg],\nonumber
\end{eqnarray}
\begin{eqnarray}\label{eq104}
\hat{P}_B(x,s;x_0)=\frac{\Xi(s)}{2s}\sqrt{\frac{v(s)}{D}}\;{\rm e}^{-(2x_N-x-x_0)\sqrt{\frac{v(s)}{D}}},
\end{eqnarray}
where $x_0<x_N$,
\begin{equation}\label{eq105}
\Xi(s)=\frac{2\lambda_1}{s\sqrt{\frac{D}{v(s)}}+\lambda_1+\lambda_2}.
\end{equation}
The functions Eqs. (\ref{eq103}) and (\ref{eq104}) coincide with Eqs. (\ref{eq45}) and (\ref{eq46}), respectively, if $\Lambda(s)\equiv\Xi(s)$. The last equation is fulfilled only when $v(s)=s$, $\lambda_1=D/\gamma_1$, and $\lambda_2=D/\gamma_2$. Thus, the boundary condition Eq. (\ref{eq52}) provides Eq. (\ref{eq102}) for the case of normal diffusion.
We mention here that the particular form of boundary condition Eq. (\ref{eq102}), namely $J=-\lambda\Delta C$ where $J$ is the particles' flux across the membrane and $\Delta C$ is the concentration difference between the membrane surfaces, is very often used, see for example \cite{aho}. The above considerations show that this boundary condition can be derived from Eq. (\ref{eq52}) assuming that there is normal diffusion without absorption with the same diffusion coefficient in both part of the system and the thin membrane is symmetrical, $\lambda_1=\lambda_2\equiv\lambda$.

	\item In \cite{kl} two models of particle random walk in a discrete homogeneous system are considered. The difference between the models is as follows. 
In the first model probability of absorption of the particle is involved in the probability density $\omega_p$ that the particle continues to exists and makes its next jump at time $t$. The particle random walk is described by the following difference equation in which the absorption term is absent
\begin{equation}\label{B1}
P_{n+1}(m;m_0)=\frac{1}{2}P_{n}(m-1;m_0)+\frac{1}{2}P_{n}(m+1;m_0).
\end{equation}	
In the second model the situation is reversed. The absorption probability is involved in the difference equation Eq. (\ref{eq1}) only, whereas particle random walk is ruled by the distribution $\omega(t)=\mathcal{L}^{-1}[1/(1+\epsilon^2 v(s)/(2\tilde{D}))]$ which is independent of an absorption coefficient. 

The second model is used in this paper because it is more useful to model diffusion in a layered system. The reason for this is that the position of functions $\hat{\omega}_{A,B}$ and parameters $D_{A,B}$ and $\kappa_{A,B}$ in the Green's functions and boundary conditions is determined by the position of probabilities $R_A$ and $R_B$. However, in order to explain why the case of the reaction $X\rightarrow \emptyset$ is not included in this model we have to use the first model which is more general than the second one. Let us assume that absorption is treated as a reaction $X+Y\rightarrow Y$, where $Y$ represents an `absorbing point'. If absorption may take place, the particle $X$ must be in the `region of absorption' generated by $Y$. The diffusing particle encounters the region with a probability $p$ after a jump. Then, absorption in the time interval $[0,t]$ may take place with probability $p(1-\rho(t))$, where $\rho(t)$ is the probability that the particle continues to exist at time $t$ when it is located in the region of absorption. If $p=1$, particle absorption may take place at any moment with constant probability regardless of particle current location, this situation corresponds to the reaction $X\rightarrow\emptyset$. 
Based on the above assumptions we get 
\begin{equation}\label{B2}
\omega_p(t)=[(1-p)+p\rho(t)]\omega(t).
\end{equation}
Let $\varphi(t)$ be the probability density that absorption takes place at the moment $t$; then, $\rho(t)=1-\int_0^t \varphi(t')dt'$. 
For $\varphi(t)=\gamma{\rm e}^{-\gamma t}$ we get (the details of the calculation are presented in the Appendix II)
\begin{eqnarray}\label{B3}
(1-p)\hat{F}(s)\left[s\hat{P}(x,s;x_0)-P(x,0;x_0)\right]\\
+p\hat{F}(s+\gamma)\left[(s+\gamma)\hat{P}(x,s;x_0)-P(x,0;x_0)\right]\nonumber\\
=\tilde{D}\frac{\partial^2 \hat{P}(x,s;x_0)}{\partial x^2},\nonumber
\end{eqnarray}
where $\hat{F}(s)=v(s)/s$. In the time domain Eq. (\ref{B3}) reads
\begin{eqnarray}\label{B4}
(1-p)\int_0^t F(t-t')\frac{\partial P(x,t';x_0)}{\partial t'}dt'\\
+p\int_0^t {\rm e}^{-\gamma(t-t')}F(t-t')\frac{\partial}{\partial t'}\left({\rm e}^{\gamma t'}P(x,t';x_0)\right)dt'\nonumber\\
=\tilde{D}\frac{\partial^2 P(x,t;x_0)}{\partial x^2}.\nonumber
\end{eqnarray}
For `classical' subdiffusion in which $v(s)=s^\alpha$, $0<\alpha<1$, we get
\begin{eqnarray}\label{B5}
(1-p)s^{\alpha-1}\left[s\hat{P}(x,s;x_0)-P(x,0;x_0)\right]\\
+p(s+\gamma)^{\alpha-1}\left[(s+\gamma)\hat{P}(x,s;x_0)-P(x,0;x_0)\right]\nonumber\\
=\tilde{D}\frac{\partial^2 \hat{P}(x,s;x_0)}{\partial x^2}.\nonumber
\end{eqnarray}
The inverse Laplace transform of Eq. (\ref{B5}) is
\begin{eqnarray}\label{B6}
(1-p)\frac{\partial^\alpha_C P(x,t;x_0)}{\partial t^\alpha}+p{\rm e}^{-\gamma t}\frac{\partial^\alpha_C \;{\rm e}^{\gamma t}P(x,t;x_0)}{\partial t^\alpha}\\
\tilde{D}\frac{\partial^2 P(x,t;x_0)}{\partial x^2},\nonumber
\end{eqnarray}
where $\partial_C^\alpha f(t)/\partial t^\alpha=(1/\Gamma(1-\alpha))\int_0^t dt' f^{(1)}(t')/(t-t')^\alpha$ is the Caputo fractional derivative of the order $\alpha$, $0<\alpha<1$. Eq. (\ref{B5}), as well as Eq. (\ref{B6}), provides qualitatively different equations for the cases of $p=1$ and $p<1$. If $p=1$ the first term on the left-hand side of Eq. (\ref{B6}) is eliminated and the equation is equivalent to the subdiffusion-reaction equation derived in \cite{sss}, see also the discussion in \cite{kl}. In this case the equation is not equivalent to Eq. (\ref{eq26}). When $p<1$ the first term gives a significant contribution to the equation. Assuming that $s\ll \gamma$, which corresponds to $t\gg 1/\gamma$, and taking into account the leading terms on the left-hand side of Eq. (\ref{B5}) we get 
\begin{eqnarray}\label{B7}
(1-p)\left[s\hat{P}(x,s;x_0)-P(x,0;x_0)\right]\\
+p\gamma^\alpha s^{1-\alpha}\hat{P}(x,s;x_0)
=\tilde{D}s^{1-\alpha}\frac{\partial^2 \hat{P}(x,s;x_0)}{\partial x^2}.\nonumber
\end{eqnarray}
Calculating the inverse Laplace transform of Eq. (\ref{B7}) we obtain Eq. (\ref{eq26}) for $D=\tilde{D}/(1-p)$ and $\kappa^2=p\gamma^\alpha/\tilde{D}$. Thus, both models provide Eq. (\ref{eq26}), the first one in the limit of long time, if absorption is treated as the reaction $X+Y\rightarrow Y$ for $p<1$. If $p=1$, which corresponds to the case of the reaction $X\rightarrow\emptyset$, the process cannot be described by Eq. (\ref{eq26}), this case is not considered in the previous sections.

The random walk model can be applied to different probability distributions $\varphi(t)$. The question arises which of the probability distributions $\varphi(t)$ is more real. The problem seems to have no unambiguous solution because the distribution is rather beyond the possibility of experimental designation. However, we can compare the concentration profiles of diffusing particles obtained as solutions to diffusion-absorption equations, which are derived from various models, with experimentally obtained concentration profiles. In this way we can check whether a model is based on the correct assumptions. Proposals for such methods based on the model considered in this paper are presented in \cite{klk,klbio}. 

The model presented in this paper assumes that absorption of diffusing particle may occur just before the next particle step, see Eqs. (\ref{eq5})--(\ref{eq7}).
We note that the discrete model is relatively easy for analytical treatment in a few cases only, when the absorption can occur just after the particle jump, just before the next jump or when the waiting time for absorption of the particle is ruled by the exponent function. In these cases, the model provides the subdiffusion-absorption equation Eq. (\ref{eq26}), which does not include the case of reaction $X\rightarrow\emptyset$. Therefore, it seems that the form of the equation depends mainly on the type of reaction that causes the disappearance of the particle. However, the influence of various functions $\varphi(t)$ on the form of diffusion-absorption equation requires further study.

	\item It may seem puzzling that in the case of $q_A\neq 0$ and $q_B\neq 0$ both probabilities $q_A$ and $q_B$, Eq. (\ref{eq55}), depend on $\epsilon$, while in the case of one-sided fully permeable membranes the non-zero probabilities, Eqs. (\ref{eq57a}) and (\ref{eq58a}), are independent of $\epsilon$. The reason is that the transition to the small parameter $\epsilon$ is done in the same way in both media $A$ and $B$. The selective properties of the membrane cannot depend on the parameter $\epsilon$ which is involved in the discrete model only. The ratio of membrane permeabilities of particles passing through the membrane in both directions $\rho(q_A,q_B)=(1-q_A(\epsilon))/(1-q_B(\epsilon))$ also must be independent of $\epsilon$. This requires that if $q_A\equiv 0$, $q_B$ is independent of $\epsilon$, and vice versa.

The frequency of particle's jumps through the thin membrane goes to infinity when $\epsilon\rightarrow 0$. If the membrane is one--sided fully permeable for the particles moving from $A$ to $B$, $q_A=0$, this property is kept for any $\epsilon$. As it is argued in Sec. \ref{SecIII}, if $q_B$ depends on $\epsilon$ then $q_B\rightarrow 1$ in the limit of small $\epsilon$; this assumption ensures that the particle will jump from $B$ to $A$ not immediately, but after a certain non-zero time. However, in this case the jump of the particle from $B$ to $A$ results in an immediate particle jump done in the opposite direction. Then, the membrane acts as a fully reflecting wall and the selective properties of the membrane are not shown in the model. 
	\item As a special case, we have considered diffusion in a system consisting of the medium $A$ in which subdiffusion without absorption occurs and the medium $B$ in which absorption of the particles occurs and mobility of particles is substantially slower comparing to the medium $A$.  Similar problem was considered in \cite{ssbs}, where normal diffusion of particles from bulk liquid phase to porous medium trough boundary film is considered; inside the porous medium particles adsorption occurs. The amount of adsorbed substance $W$ evolves over time according to the formula $W(t)=\beta\sqrt{t}+\lambda$, where $\beta$ and $\lambda$ are constants which physical meaning is discussed in the above cited paper. We note that this function is obtained from Eq. (\ref{eq85}) in the long time limit putting $\alpha_A=1$, $\beta=2C_0\sqrt{D_A}/\sqrt{\pi}$, and $\lambda=-\gamma_A/\kappa_B$.
	
We note that the observation of the temporal evolution of the amount of substance in the medium $A$ shows the properties of the process taking place in the medium $B$. This is of practical significance in the case when observation of diffusion in the medium $B$ is not possible, see the discussion in \cite{klk}.
\end{enumerate}

\subsection*{Suggestion: how to include superdiffusion into the model}

In the model the kind of diffusion is defined by the function $v$ which controls the time which is needed to take particle next step. The function well defines normal diffusion, subdiffusion and slow subdiffusion. If we were able to define superdiffusion by the function $v$, this process could be included in the model. In such a case, it would be possible, for example, to determine the Green's function and boundary conditions for a system consisting of a subdiffusive medium and superdiffusive one separated by a thin membrane. The main problem is that superdiffusion is defined in the CTRW model as a random walk process for which the mean time which is needed to take particle next step $\left\langle \tau\right\rangle$ is finite whereas the length of a particle jump, which is a random variable, has an infinite variance. These assumptions lead to the superdiffusion equation with the Riesz fractional derivative with respect to the spatial variable \cite{mk}. This process is characterized by the following relation
\begin{equation}\label{eq91}
\left\langle \left(\Delta x\right)^2(t)\right\rangle\sim t^\beta\;,\;1<\beta<2.
\end{equation}
This relation is often treated as the definition of superdiffusion. Due to Eq. (\ref{eq24a}), putting 
\begin{equation}\label{eq92}
v(s)=s^{\beta}\;,
\end{equation} 
we get $\left\langle \left(\Delta x\right)^2(t)\right\rangle=2Dt^{\beta}/\Gamma(1+\beta)$. Thus, the relation (\ref{eq91}) is fulfilled supposing $1<\beta<2$. However, the interpretation of the diffusion process generated by Eq. (\ref{eq92}) is not obvious if $\beta>1$. Let us consider two functions characterizing the random walk of the particle, namely $\left\langle \tau\right\rangle$ and the frequency of particle jumps between neighbouring sites $\nu(t)$. From Eqs. (\ref{eq24a}), (\ref{eq53a}), and (\ref{eq92}) we get
\begin{eqnarray}\label{eq93}
\left\langle \tau\right\rangle&=&\left\{\begin{array}{ll}
      0\;,&\beta>1\;,\\
			\tau_0\neq 0\;,&\beta=1\;,\\
      \infty\;,&0<\beta<1\;,
    \end{array}\right.
\end{eqnarray}
and
\begin{equation}\label{eq94}
\nu(t)=\frac{2Dt^{\beta-1}}{\epsilon^2 \Gamma(\beta)}.
\end{equation}
Then, defining superdiffusion by Eq. (\ref{eq92}) with $1<\beta<2$, we get from Eq. (\ref{eq93}) that the average waiting time for the particle next step is equal to zero. However, from Eq. (\ref{eq94}) we obtain that the frequency of the particle steps is anomalously large and goes to infinity in the limit of long time even for non-zero $\epsilon$. Thus, the interpretation is that extreme high frequency of particle's steps leads to $\left\langle \tau\right\rangle =0$. Such an interpretation may be considered controversial. Moreover, it is also not clear if Eq. (\ref{eq91}) alone defines superdiffusion, see the discussion in \cite{dybiec}. In a lot of physical models the assumptions that simplify considerations but which interpretation is not obvious are made. However, such models can be useful and provide the results confirmed experimentally. The problem of whether superdiffusion can be included in the model using the function $v$ Eq. (\ref{eq92}) with $\beta>1$ requires further considerations which will be presented elsewhere.

\section*{Acknowledgements}

The author thanks Eli Barkai for helpful discussion.

\section*{Appendix I. How to calculate the inverse Laplace transform}

One of the main problems in the presented model is the calculation of the inverse Laplace transform. For the case of classic subdiffusion, the following equation is useful
\begin{eqnarray}\label{eqa1}
\mathcal{L}^{-1}\left[s^\nu {\rm e}^{-as^\beta}\right]\equiv f_{\nu,\beta}(t;a)\\
=\frac{1}{t^{\nu+1}}\sum_{k=0}^\infty{\frac{1}{k!\Gamma(-k\beta-\nu)}\left(-\frac{a}{t^\beta}\right)^k}\;,\nonumber
\end{eqnarray}
$a,\beta>0$; the function $f_{\nu,\beta}$ is a special case of the Wright function and the H-Fox function.
The following theorem can be also useful \cite{krylov}
{\it Let $\hat{g}(s)\rightarrow 0$ when $s\rightarrow 0$, ${\rm Re}\; s<c$, $c$ is a positive real number, and $\hat{g}$ does not have any singularities except the point $s=0$ which is a branch point. Then, if 
\begin{equation}\label{eqa2}
\hat{g}(s)=s^\alpha\sum_{n=0}^\infty a_n s^{n\beta},
\end{equation}
where $\beta>0$, then
\begin{equation}\label{eqa3}
g(t)=\frac{1}{t^{\alpha+1}}\sum_{n=0}^\infty \frac{a_n}{\Gamma(-\alpha-n\beta)}\frac{1}{t^{n\beta}},
\end{equation}
and in Eq. (\ref{eqa3}) all terms for which $\alpha+n\beta$ is a natural number are omitted.}
If the inverse Laplace transforms cannot be calculated using standard formulas, we use the series expansion of the transform with respect to $s$. For a typical situation we have $\hat{f}(s){\rm e}^{-bs^\alpha}={\rm e}^{-bs^\alpha}\sum_{n=0}^\infty a_n s^{\beta+n\nu}$, $\alpha,b>0$, and we calculate the inverse Laplace transform term by term using Eq. (\ref{eqa1}). The functions obtained in such way can be considered in the limit of small $s$, which corresponds to the limit of long time. Then, the first few terms of both series Eqs. (\ref{eqa2}) and (\ref{eqa3}) can be taken into account. To roughly estimate the time for which obtained functions are correct we assume that the last term of the reduced series labelled by $n$ is much larger that the next one. Applying this rule to both series Eqs. (\ref{eqa2}) and (\ref{eqa3}), and putting $n=1$, $a=|a_1/a_2|$ and $\alpha=1$ we get 
\begin{equation}\label{eqa4}
s^\beta\ll a\Leftrightarrow t^{\beta}\gg\frac{|\Gamma(-1-\beta)|}{a|\Gamma(-1-2\beta)|}.
\end{equation}

To calculate the inverse Laplace transform of the Green's functions for slow subdiffusion the following Strong Tauberian Theorem can been used \cite{hughes}
{\it If $\phi(t)\geq 0$, $\phi(t)$ is ultimately monotonic like $t\longrightarrow\infty$, $\mathcal{R}$ is slowly--varying at infinity and $0<\rho<\infty$, then each of the relations 
\begin{equation}\label{eqa5}
\hat{\phi}(s)\approx \frac{\mathcal{R}(1/s)}{s^\rho}
\end{equation}
as $s\longrightarrow 0$ and
\begin{equation}\label{eqa6}
\phi(t)\approx \frac{\mathcal{R}(t)}{\Gamma(\rho)t^{1-\rho}}
\end{equation}
as $t\longrightarrow\infty$ implies the other}.

The following formula is also helpful when analysing the slow subdiffusion process
\begin{equation}\label{eqa7}
\mathcal{L}^{-1}\left[\hat{f}\left(\frac{1}{s}\right)\hat{g}(s)\right]=f(t)g(t),
\end{equation}
where $s\rightarrow 0$, $t\rightarrow\infty$, and $f$ is a slowly varying function. The `heuristic' derivation of this formula is as follows. Since for slowly varying function there is $f(u/s)\approx f(1/s)$ when $s\rightarrow 0$, we have $\int_0^\infty {\rm e}^{-st}f(t)g(t)dt=\int_0^\infty {\rm e}^{-u}f(u/s)g(u/s)du/s=f(1/s)\int_0^\infty {\rm e}^{-st}g(t)dt$.

\section*{Appendix II. Derivation of Eqs. (\ref{B3})--(\ref{B7})}

The generating function of Eq. (\ref{B1}) is given by Eqs. (\ref{eq11}) and (\ref{eq12}) for $R=0$. Taking into account Eq. (\ref{B2}), the probability that the particle does not perform any step in the time interval $[0,t]$ and continues to exists at time $t$ reads
\begin{equation}\label{B8}
U_p(t)=\left[1-p(1-\rho(t))\right]\left[1-\int_0^t \omega(t')dt'\right].
\end{equation}
The Laplace transform of probability $P(m,t;m_0)$ is
\begin{equation}\label{B9}
\hat{P}(m,s;m_0)=\hat{U}_p(s)S(m,\hat{\omega}_p(s);m_0).
\end{equation} 
Assuming $\varphi(t)=\gamma{\rm e}^{-\gamma t}$ we get $\rho(t)={\rm e}^{-\gamma t}$. Then, from Eq. (\ref{B2}) we get
\begin{equation}\label{B10}
\hat{\omega}_p(s)=(1-p)\hat{\omega}(s)+p\hat{\omega}(s+\gamma).
\end{equation}
Supposing $\hat{\omega}(s)=1/(1+\epsilon^2 v(s)/(2\tilde{D}))$ we get for small $\epsilon$
\begin{equation}\label{B11}
\hat{\omega}_p(s)=1-\frac{\epsilon^2}{2\tilde{D}}\left[(1-p)v(s)+pv(s+\gamma)\right],
\end{equation}
and
\begin{equation}\label{B12}
\hat{U}_p(s)=\frac{\epsilon^2}{2\tilde{D}}\left[(1-p)\frac{v(s)}{s}+p\frac{v(s+\gamma)}{s+\gamma}\right].
\end{equation}
From Eq. (\ref{eq17}), in which $R=0$, and Eqs. (\ref{eq8}), (\ref{eq9}), and (\ref{B9})--(\ref{B12}) we obtain Eq. (\ref{B3}). Next, due to the relation 
\begin{eqnarray}\label{B13}
\mathcal{L}^{-1}\left[F(s+\gamma)\left((s+\gamma)\hat{f}(s)-f(0)\right)\right]\\
=\int_0^t {\rm e}^{-\gamma(t-t')}F(t-t')\frac{d}{dt'}\left({\rm e}^{\gamma t'}f(t')\right)dt',\nonumber
\end{eqnarray}
we get Eq. (\ref{B4}). Assuming $v(s)=s^\alpha$, $0<\alpha<1$, we get Eq. (\ref{B5}). Due to the equation
\begin{equation}\label{B14}
\frac{\partial^\alpha_C f(t)}{\partial t^\alpha}=\mathcal{L}^{-1}\left[s^\alpha\hat{f}(s)-s^{\alpha-1}f(0)\right],
\end{equation}
and Eq. (\ref{B13}) we obtain Eq. (\ref{B6}).
To derive Eq. (\ref{B7}) from Eq. (\ref{B5}) we use the approximation $(\gamma+s)^{\beta}\approx\gamma^\beta(1+\beta s/\gamma)$ under assumption that $s\ll\gamma$.

\end{document}